\documentclass[a4paper]{amsart}
\usepackage{epic,eepic,amsmath,amsthm,amssymb}
\usepackage{epsfig,cite,indentfirst,oldgerm}
\usepackage{multicol,boxedminipage}

\begin{document}

\title[Phase Model and Toda Hierarchy]
{Phase model expectation values and the 
2-Toda hierarchy}

\author{M. Zuparic}

\address{Department of Mathematics and Statistics,
         University of Melbourne, 
         Parkville, Victoria 3010, Australia.}
\email{mzup@ms.unimelb.edu.au}

\date{}

\newcommand{\field}[1]{\mathbb{#1}}
\newcommand{\C}{\field{C}}
\newcommand{\N}{\field{N}}
\newcommand{\Z}{\field{Z}}
\newcommand{\R}{\field{R}}

\begin{abstract}
We show that the scalar product of the phase model on a 
finite rectangular lattice is a (restricted) $\tau$-function
of the 2-Toda hierarchy. Using this equivalence we then show
that the wave-functions of the hierarchy correspond to 
certain classes of boundary correlation functions of the model.
\end{abstract}

\maketitle
\newtheorem{ca}{Figure}
\newtheorem{corollary}{Corollary}
\newtheorem{definition}{Definition}
\newtheorem{example}{Example}
\newtheorem{lemma}{Lemma}
\newtheorem{notation}{Notation}
\newtheorem{proposition}{Proposition}
\newtheorem{remark}{Remark}
\newtheorem{condition}{Condition}
\newtheorem{theorem}{Theorem}

\hyphenation{boson-ic
             ferm-ion-ic
             para-ferm-ion-ic
             two-dim-ension-al
             two-dim-ension-al
             rep-resent-ative
             par-tition
             anti-comm-uta-tion}

\newtheorem{Theorem}{Theorem}[section]
\newtheorem{Corollary}[Theorem]{Corollary}
\newtheorem{Proposition}[Theorem]{Proposition}
\newtheorem{Conjecture}[Theorem]{Conjecture}
\newtheorem{Lemma}[Theorem]{Lemma}
\newtheorem{Example}[Theorem]{Example}
\newtheorem{Note}[Theorem]{Note}
\newtheorem{Definition}[Theorem]{Definition}
                                                                               
\renewcommand{\mod}{\textup{mod}\,}
\newcommand{\wt}{\text{wt}\,}

\newcommand{\T}{{\mathcal T}}
\newcommand{\U}{{\mathcal U}}
\newcommand{\tT}{\tilde{\mathcal T}}
\newcommand{\tU}{\widetilde{\mathcal U}}
\newcommand{\Y}{{\mathcal Y}}
\newcommand{\B}{{\mathcal B}}
\newcommand{\D}{{\mathcal D}}
\newcommand{\M}{{\mathcal M}}
\renewcommand{\P}{{\mathcal P}}

\hyphenation{And-rews
             Gor-don
             boson-ic
             ferm-ion-ic
             para-ferm-ion-ic
             two-dim-ension-al
             two-dim-ension-al}

\setcounter{section}{-1}

\section{Introduction}\label{introduction}
In \cite{DWPF}, it was observed that the $N\times N$ 
domain wall partition function, $Z_N$, of the six vertex 
model is, up to a multiplicative factor, a 
$\tau$-function of the KP hierarchy. In \cite{XXZ}, the length $M$ XXZ spin-$\frac{1}{2}$ chain 
and its associated scalar product, $\langle \{ \lambda \}|
\{\mu \} \rangle$, were considered. Restricting either the initial or the final state
to a Bethe eigenstate, the resulting expression
is again a KP $\tau$-function. 

In this work we further extend the known correspondences
between integrable quantum lattice models and classical 
hierarchies of non linear partial differential equations 
in the following way\footnote{We refer to \cite{purplebook}
for an introduction to correspondences between integrable quantum models and differential equations.}. We show that the scalar product of 
the phase model \cite{Bog1,Bog2}, up to a multiplicative factor, is a (restricted) $\tau$-function of the
2-Toda hierarchy \cite{Toda1,Toda2}, where the Toda time variables
are power sums of the rapidities. We then consider the two 
types of wave-functions from the Toda theory, 
$\hat{w}^{(\infty)}$ and $\hat{w}^{(0)}$, and show that they
correspond to a certain class of boundary correlation function 
from the phase model perspective. We additionally give a 
single determinant form for each of these correlation
functions.

In section $\mathbf{1}$, we recall known results about the 
finite 2-Toda hierarchy, its construction from the wave-matrix
\textit{initial value problem} and the 
$\tau$-function as a finite bilinear sum of character/Schur polynomials. In $\mathbf{2}$, we introduce the phase model and the aforementioned lattice model/hierarchy correspondence. Additionally we use known combinatorial bijections to express the state vectors of the model as weighted sums of various objects. In $\mathbf{3}$, we use the weighted sum expressions of the state vectors to show that the Toda wave-functions correspond to specific classes of boundary correlation functions and give their single determinant form. In $\mathbf{4}$, we offer some remarks.
\section{The finite 2-Toda hierarchy}
\subsection{Definition of the hierarchy.} The details of this section can mostly be found in \cite{Toda1,Toda2}. We begin by giving the definition of the 2-Toda hierarchy, with two sets of $(n-m-1)$ time variables, in terms of four distinct Lax type systems of first order differential equations.\\
\indent Defining the following shift matrices,
\small\begin{equation*}
 \Lambda^{\pm j}_{[m,n)} \equiv (\delta_{k \pm j,l})_{k,l \in \{m,\dots,n-1 \}}
\end{equation*}\normalsize
the general matrix $A \in gl(n-m)$ is written in the form,
\small\begin{equation*}
A= \sum^{n-m-1}_{j=-n+m+1}a_j(s)\Lambda^j_{[m,n)}
\end{equation*}\normalsize
where $m \le s \le n-1 $ denotes the row of the matrix $A$. $A$ is said to be a strictly lower triangular, $A=(A)_-$, if $a_j(s)=0$ for $j \ge 0$ and upper triangular, $A=(A)_+$, if $a_j(s)=0$ for $j < 0$,
\small\begin{equation*}
(A)_+ = \sum^{n-m-1}_{j=0} a_j(s)\Lambda^j_{[m,n)} \textrm{  ,  } (A)_- = \sum^{-1}_{j=-n+m+1} a_j(s)\Lambda^j_{[m,n)}
\end{equation*}\normalsize
We define two sets of time flows $\vec{x}$ and $\vec{y}$ as,
\small\begin{equation*}
\vec{x} = \{ x_1,x_2,\dots, x_{n-m-1}\} \textrm{  ,  } \vec{y} = \{ y_1,y_2,\dots, y_{n-m-1}\}
\end{equation*}\normalsize
and introduce $L(\vec{x},\vec{y}), M(\vec{x},\vec{y}), B_k(\vec{x},\vec{y}), C_k(\vec{x},\vec{y}) \in gl(n-m)$ where,
\small\begin{equation*}\begin{array}{lll}
\displaystyle L = \sum^{1}_{j=-n+m+1} b_j(s,\vec{x},\vec{y}) \Lambda^j_{[m,n)}, & b_1(s) = 1,&  B_k = (L^k)_+\\
\displaystyle M = \sum^{n-m-1}_{j=-1} c_j(s,\vec{x},\vec{y}) \Lambda^j_{[m,n)}, & c_{-1}(s) \ne 0, & C_k = (M^k)_- 
\end{array}\end{equation*}\normalsize
We define the 2-Toda hierarchy as the following Lax type system of differential equations,
\small\begin{equation}
\partial_{x_k} L = [B_k,L] \textrm{  ,  } \partial_{y_k} L = [C_k,L] \textrm{  ,  } \partial_{x_k} M = [B_k,M]\textrm{  ,  }  \partial_{y_k} M = [C_k,M]
\label{lax}\end{equation}\normalsize
or equivalently (theorem 1.1 of \cite{Toda1}), the Zakharov-Shabat system,
\small\begin{equation}\begin{split}
\partial_{x_{j}} B_{k}-\partial_{x_{k}} B_{j} + \left[ B_{k},B_{j} \right] \textrm{  $=$  }0 &\textrm{  ,   } \partial_{y_{j}} C_{k}-\partial_{y_{k}} C_{j} + \left[ C_{k},C_{j} \right] \textrm{  $=$  } 0\\
\partial_{y_{j}} B_{k}-\partial_{x_{k}} C_{j} + \left[ B_{k},C_{j} \right] \textrm{  $=$  } 0
\end{split}\label{linear2}\end{equation}\normalsize
\textbf{Compatibility conditions.} It can be shown that the above systems are the compatibility conditions of the linear operator equations,
\small\begin{equation}
\partial_{x_{j}} W^{(\infty/0)} = B_j  W^{(\infty/0)} \textrm{  ,  } \partial_{y_{j}} W^{(\infty/0)} = C_j W^{(\infty/0)}
\label{linear3}\end{equation}\normalsize
where $W^{(\infty/0)}=W^{(\infty/0)}(\vec{x},\vec{y}) \in GL(n-m)$ are referred to as \textit{wave-matrices}.
\subsection{The initial value problem.} By defining the constant matrix $A \in GL(n-m) = (a_{ij})_{i,j = m, \dots,n-1}$, such that $\textrm{det}\left[ a_{ij} \right]_{i,j=m\dots,s-1} \ne 0$, $m < s \le n$, it is possible to find wave-matrices, $W^{(\infty)}$ and $W^{(0)}$, such that
\small\begin{equation}
W^{(0)}=W^{(\infty)} A
\label{1}\end{equation}\normalsize
where $W^{(\infty)}$ and $W^{(0)}$ have the specific form
\small\begin{equation*} 
\begin{split}
W^{(\infty)} = \hat{W}^{(\infty)} \exp \left[\sum^{n-m-1}_{k = 1} x_{k} \Lambda^{k}_{[m,n)}\right] \textrm{  ,  }\hat{W}^{(\infty)} = \left(\hat{w}^{(\infty)}_{i-j}(i,\vec{x},\vec{y})\right)_{m \le i,j \le n-1}  \\
W^{(0)} = \hat{W}^{(0)} \exp \left[\sum^{n-m-1}_{k = 1} y_{k} (\Lambda^T_{[m,n)})^{k}\right] \textrm{  ,  }  \hat{W}^{(0)} = \left(\hat{w}^{(0)}_{j-i}(i,\vec{x},\vec{y})\right)_{m \le i,j \le n-1}
\end{split}
\end{equation*}\normalsize
\small\begin{equation}
\begin{split}
 \hat{w}^{(\infty)}_j = \left\{ \begin{array}{cc} 
0 & j < 0\\
1 & j =0 \end{array} \right. \textrm{  ,  }  \hat{w}^{(0)}_j =  \left\{ \begin{array}{cc}
0 & j < 0\\ 
 \hat{w}^{(0)}_j(\vec{x},\vec{y}) \ne const. & j =0
\end{array}\right.
\end{split}
\label{ogle}\end{equation}\normalsize
\noindent The remaining non zero entries of the wave-matrices, $\hat{W}^{(\infty)}$ and $\hat{W}^{(0)}$, are given by (proposition 3.1 of \cite{Toda2}),
\small\begin{equation}\begin{split}
\hat{w}^{(\infty)}_{k}(s,\vec{x},\vec{y}) =& (-1)^k\frac{\textrm{det}\left[a_{ij}(\vec{x},\vec{y}) \right]_{\genfrac{}{}{0mm}{}{i=m,\dots,\hat{s-k},\dots ,s}{j=m,\dots,s-1}}}{\textrm{det}\left[a_{ij}(\vec{x},\vec{y}) \right]_{i,j=m,\dots,s-1}}, \textrm{  for  } \left\{ \begin{array}{c} 0 \le k \le s-m\\ m < s \le n-1\end{array}\right. \\
\hat{w}^{(0)}_{k}(s,\vec{x},\vec{y}) =& \frac{\textrm{det}\left[a_{ij}(\vec{x},\vec{y}) \right]_{\genfrac{}{}{0mm}{}{i=m,\dots,s}{j=m,\dots,s-1,s+k}}}{\textrm{det}\left[a_{ij}(\vec{x},\vec{y}) \right]_{i,j=m,\dots,s-1}}, \textrm{  for  } \left\{ \begin{array}{c}  0 \le k \le n-s-1\\ m < s \le n-1\end{array}\right.
\end{split}
\label{wavemat1}\end{equation}\normalsize
where,
\small\begin{equation*}\begin{split}
\left( a_{ij}(\vec{x},\vec{y}) \right)^{}_{m \le i,j \le n-1} = \exp \left[\sum^{n-m-1}_{k = 1} x_{k} \Lambda^{k}_{[m,n)}\right] A \exp \left[-\sum^{n-m-1}_{k = 1} y_{k} (\Lambda^T_{[m,n)})^{k}\right] 
\end{split}
\end{equation*}\normalsize
and $s$ refers to the row of the entry. The entries of the inverse of the wave-matrices $\left(\hat{W}^{(0)}(\vec{x},\vec{y})\right)^{-1}$ and $\left(\hat{W}^{(\infty)}(\vec{x},\vec{y})\right)^{-1}$ are similarly given as
\small\begin{equation}\begin{split}
\hat{w}^{*(0)}_{k}(s,\vec{x},\vec{y})& = (-1)^k\frac{\textrm{det}\left[a_{ij}(\vec{x},\vec{y}) \right]_{\genfrac{}{}{0mm}{}{i=m,\dots,s-1}{j=m,\dots,\hat{s-k},\dots,s}}}{\textrm{det}\left[a_{ij}(\vec{x},\vec{y}) \right]_{i,j=m,\dots,s}},\textrm{  for  }\left\{ \begin{array}{c} 0 \le k \le s-m \\ m < s \le n-1 \end{array}\right. \\
\hat{w}^{*(\infty)}_{k}(s,\vec{x},\vec{y}) &= \frac{\textrm{det}\left[a_{ij}(\vec{x},\vec{y}) \right]_{\genfrac{}{}{0mm}{}{i=m,\dots,s-1,s+k}{j=m,\dots,s}}}{\textrm{det}\left[a_{ij}(\vec{x},\vec{y}) \right]_{i,j=m,\dots,s}},\textrm{  for  } \left\{ \begin{array}{c}  0 \le k \le n-s-1\\ m < s \le n-1\end{array}\right.
\end{split}
\label{wavemat2}\end{equation}\normalsize
where $s$ now refers to the column of the entry, as opposed to the row.\\
\\
\textbf{The generalized Lax and Zakharov-Shabat systems.} From proposition 3.2 in \cite{Toda2}, the following matrices,
\small\begin{equation*}
L = W^{(\infty)} \Lambda_{[m,n)} \left(  W^{(\infty)}\right)^{-1} \textrm{  ,  } M = W^{(0)}\Lambda^T_{[m,n)} \left(  W^{(0)} \right)^{-1}  
\end{equation*}\normalsize
and $B_{k} = \left\{ L^{k} \right\}_+ \textrm{  ,  } C_{k} = \left\{ M^{k} \right\}_-$, satisfy the linear operator equations (eq. \ref{linear3}), the Zakharov-Shabat equations (eq. \ref{linear2}) and the Lax equations (eq. \ref{lax}) which define the 2-Toda hierarchy.
\subsection{Tau-function of the 2-Toda hierarchy.} The $\tau$-function, $\tau(s,\vec{x},\vec{y})$, is a function of the time parameters $\vec{x}$ and $\vec{y}$ and an additional parameter, $s$, which corresponds to the row number of $W^{(\infty)},W^{(0)}$ or the column number of $\left( W^{(\infty)} \right)^{-1}$, $\left( W^{(0)} \right)^{-1}$. The derivatives of $\tau(s,\vec{x},\vec{y})$ correspond to the entries of the wave-matrices and using this fact, we can express the 2-Toda hierarchy in a single integral bilinear form. 
\begin{proposition} For the function,
\small\begin{equation}
\tau(s,\vec{x},\vec{y}) = \textrm{det}\left[a_{ij}(\vec{x},\vec{y}) \right]_{m \le i,j \le s-1}
\label{5}\end{equation}\normalsize
the following four relations hold,
\small\begin{equation}\begin{array}{ll}
\displaystyle \hat{w}^{(\infty)}_{k}(s) = \frac{\zeta_k \left(-\tilde{\partial}_{\vec{x}}\right)\tau(s,\vec{x},\vec{y})}{\tau(s,\vec{x},\vec{y})} &\displaystyle \hat{w}^{(0)}_{k}(s) = \frac{\zeta_k \left(-\tilde{\partial}_{\vec{y}}\right)\tau(s+1,\vec{x},\vec{y})}{\tau(s,\vec{x},\vec{y})} \\
\displaystyle \hat{w}^{*(\infty)}_{k}(s) = \frac{\zeta_k \left(\tilde{\partial}_{\vec{x}} \right)\tau(s+1,\vec{x},\vec{y})}{\tau(s+1,\vec{x},\vec{y})} &\displaystyle \hat{w}^{*(0)}_{k}(s) = \frac{\zeta_k \left(\tilde{\partial}_{\vec{y}} \right)\tau(s,\vec{x},\vec{y})}{\tau(s+1,\vec{x},\vec{y})}
\end{array}\label{ned}\end{equation}\normalsize
where,
\small\begin{equation*}\begin{split}
\tilde{\partial}_{\vec{x}} = \left(\partial_{x_1},  \frac{1}{2}\partial_{x_2}, \frac{1}{3}\partial_{x_3}, \dots \right) \textrm{  ,  } \tilde{\partial}_{\vec{y}} =  \left(\partial_{y_1},  \frac{1}{2}\partial_{y_2}, \frac{1}{3}\partial_{y_3}, \dots \right)
\end{split}\end{equation*}\normalsize
and the generating function for the \textbf{one row character polynomial}, $\zeta_k(\vec{x})$, is given by,
\small\begin{equation}\begin{split}
\sum^{\infty}_{k=0} z^k \zeta_k(\vec{x}) &= \exp \left\{ \sum^{n-m-1}_{j=1}z^j x_j \right\}
\end{split}\end{equation}\normalsize
\end{proposition}
\textbf{Proof.} If the above four relations are true then their weighted summations are given by, 
\small\begin{equation}\begin{array}{ll}
\displaystyle \sum^{s-m}_{k=0} \lambda^k \hat{w}^{(\infty)}_{k}(s) = \frac{\tau(s,\vec{x}-\vec{\epsilon}(\lambda),\vec{y})}{\tau(s,\vec{x},\vec{y})} &\displaystyle \sum^{n-s-1}_{k=0} \lambda^k \hat{w}^{(0)}_{k}(s) = \frac{\tau(s+1,\vec{x},\vec{y}-\vec{\epsilon}(\lambda))}{\tau(s,\vec{x},\vec{y})}\\
\displaystyle \sum^{n-s-1}_{k=0} \lambda^k \hat{w}^{*(\infty)}_{k}(s) = \frac{\tau(s+1,\vec{x}+\vec{\epsilon}(\lambda),\vec{y})}{\tau(s+1,\vec{x},\vec{y})} &\displaystyle \sum^{s-m}_{k=0} \lambda^k \hat{w}^{*(0)}_{k}(s) =  \frac{\tau(s,\vec{x},\vec{y}+\vec{\epsilon}(\lambda))}{\tau(s+1,\vec{x},\vec{y})}\label{H.20}
\end{array}\end{equation}\normalsize
where, $\vec{\epsilon}(\lambda) = (\lambda, \frac{\lambda^2 }{2},\frac{\lambda^3 }{3}, \dots )$. By using the methods in proposition 3.4 of \cite{Toda2} we explicitly obtain,
\small\begin{equation*}\begin{split}
\tau(s,\vec{x} \mp \vec{\epsilon}(\lambda),\vec{y}) = \textrm{det}\left[ (1-\lambda \Lambda_{[m,n)})^{\pm 1} A(\vec{x},\vec{y}) \right]_{m \le i,j \le s-1}\\
\tau(s,\vec{x} ,\vec{y}\mp \vec{\epsilon}(\lambda)) = \textrm{det}\left[  A(\vec{x},\vec{y}) (1-\lambda \Lambda^T_{[m,n)})^{\mp 1} \right]_{m \le i,j \le s-1}
\end{split}\end{equation*}\normalsize
which upon expanding as a polynomial in $\lambda$ we obtain the required summations in eq. \ref{H.20}. $\square$\\
\\
\textbf{2-Toda Bilinear relation.} The function $\tau(s,\vec{x},\vec{y})$ defined in eq. \ref{5} satisfies the following bilinear relation,
\small\begin{equation}\begin{split}
 \oint \frac{d \lambda}{2 \pi i} \lambda^{s'-s-2} \exp \left\{ \sum^{n-m-1}_{l=1}(y_{l}-y'_{l}) \lambda^{l} \right\} \frac{\tau\left(s+1,\vec{x},\vec{y}-\vec{\epsilon}\left(\frac{1}{\lambda}\right) \right)}{\tau(s,\vec{x},\vec{y})}\frac{\tau \left(s'-1,\vec{x}',\vec{y}'+\vec{\epsilon}\left(\frac{1}{\lambda} \right)\right)}{\tau\left(s',\vec{x}',\vec{y}'\right)}\\
= \oint \frac{d \lambda}{2 \pi i} \lambda^{s-s'} \exp \left\{ \sum^{n-m-1}_{l=1}(x_{l}-x'_{l}) \lambda^{l} \right\} \frac{\tau\left(s,\vec{x}-\vec{\epsilon}\left(\frac{1}{\lambda}\right),\vec{y}\right)}{ \tau\left(s,\vec{x},\vec{y}\right)}\frac{\tau\left(s',\vec{x}'+\vec{\epsilon}\left(\frac{1}{\lambda}\right),\vec{y}'\right)}{\tau\left(s',\vec{x}',\vec{y}'\right)}
\label{biline}\end{split}\end{equation}\normalsize
for general $s,s',\vec{x},\vec{x}',\vec{y},\vec{y}'$. The integration $\oint \frac{d \lambda}{2 \pi i}$ simply refers to the algebraic operation of obtaining the coefficient of $\frac{1}{\lambda}$.\\
\\
\textbf{Polynomial expressions of the $\tau$-function.} Rewriting the shift matrix exponentials appropriately,
\small\begin{equation*}\begin{split}
\exp\left\{ \sum^{n-m-1}_{l=1}x_{l}\Lambda^{l}_{[m,n)} \right\} &= \sum^{n-m-1}_{j=0}\zeta_j (\vec{x}) \Lambda^{j}_{[m,n)}= \left( \zeta_{j-i}(\vec{x}) \right)_{m \le i,j \le n-1}\\
\exp\left\{- \sum^{n-m-1}_{l=1}y_{l}\left(\Lambda^{T}_{[m,n)}\right)^{l} \right\} &= \sum^{n-m-1}_{j=0}\zeta_j (-\vec{y}) \left(\Lambda^{T}_{[m,n)}\right)^j= \left( \zeta_{i-j}(-\vec{y}) \right)_{m \le i,j \le n-1}
\end{split}\end{equation*}\normalsize
and using the repeated application of the Cauchy-Binet identity, the $\tau$-function becomes,
\small\begin{equation}
\tau\left(s,\vec{x},\vec{y}\right) = \sum_{\{ \lambda\}\{ \mu\} \subseteq (n-s)^{(s-m)}}A_{\{ \lambda\}\{ \mu\}} \chi_{\{ \lambda\}}(\vec{x}) \chi_{\{ \mu\}}(-\vec{y}) \label{H.99}
\end{equation}\normalsize
where $\{ \lambda\}$ and $\{ \mu\}$ are partitions contained within the box of dimensions $(n-s)^{(s-m)}$, $\chi_{\{ \lambda\}}(\vec{x})$ is the character polynomial given by,
\small\begin{equation}
\chi_{\{ \lambda\}}(\vec{x}) = \textrm{det} \left[ \zeta_{ \lambda_{i}+j-i}(\vec{x})\right]_{1 \le i,j \le s-m}
\end{equation}\normalsize
and,
\small\begin{equation*}
A_{\{\lambda\}\{\mu\}} = \textrm{det} \left[   a_{\lambda_{s-m+1-i}+i +m-1 ,\mu_{s-m+1-j}+j+ m-1} \right]_{1 \le i,j \le s-m}
\end{equation*}\normalsize
\subsection{Restricting the time variables.} In order to make contact with the phase model, it is necessary to restrict the time variables in such a way that the $\tau$-function becomes an element of the symmetric polynomial ring,\\$\field{C}\{[u_1,\dots,u_{s-m}]^{S_{s-m}},[v_1,\dots,v_{s-m}]^{S_{s-m}}\}$. \\
\indent In the remainder of this work we shall use the convention that,
\begin{itemize}
\item{$\tau(\vec{x},\vec{y})$ denotes that the time variables are algebraically independent.}
\item{$\tau(\vec{u},\vec{v})$ denotes that the time variables are algebraically dependent, and $\tau(\vec{u},\vec{v})$ is an element of the aforementioned symmetric polynomial ring. We shall refer to $\tau(\vec{u},\vec{v})$ as a \textbf{restricted} $\tau$-function.}\\
\end{itemize}
\textbf{Schur polynomials.} Changing from time parameters to symmetric power sums\footnote{We define the symmetric power sum of order $k$ as, $p_k(\vec{u}) = \sum^{s-m}_{j=1} u^k_j$.},
\small\begin{equation}
x_{k} = \frac{1}{k}p_k(u_1,\dots,u_{s-m}) \textrm{  ,  } - y_{k} = \frac{1}{k} p_k(v_1,\dots,v_{s-m}) \textrm{  ,  }  k \in \{1,\dots, n-m-1\}
\label{MIWA}\end{equation}\normalsize
the one row character polynomials, $\zeta_i(\vec{x})$ and $\zeta_i(-\vec{y})$, become complete homogeneous symmetric polynomials\footnote{The complete homogeneous symmetric polynomials, $h_k(\vec{u})$, are generated by $\sum^{\infty}_{k=0}z^k h_k(\vec{u}) = \prod^{s-m}_{j=1} \frac{1}{1-z u_j} = \exp \left\{ \sum^{\infty}_{j=1} z^j \frac{1}{j} p_j(\vec{u}) \right\}$. For additional information see section I.2 of \cite{MacD}.},
\begin{eqnarray*}
\zeta_i(\vec{x}) \rightarrow h_i (u_1,\dots,u_{s-m}) &,&\zeta_i(-\vec{y}) \rightarrow h_i (v_1,\dots,v_{s-m}) 
\end{eqnarray*}
Hence the character polynomials in the $\tau$-function become Schur polynomials,
\small\begin{equation}\begin{split}
\tau\left(s,\vec{u},\vec{v}\right) &=   \sum_{\{\lambda\}\{\mu\} \subseteq (n-s)^{(s-m)}}A_{\{\lambda\}\{\mu\}} S_{\{\lambda\}}(\vec{u}) S_{\{\mu\}}(\vec{v})
\end{split}\label{restrict}\end{equation}\normalsize
\section{The phase model}
Most of the material in this section can be found in \cite{Bog1,Bog2}. Consider the bosonic algebra\footnote{We note that this algebra is the $q = 0$ limit of the $q$-boson algebra.} generated by the three operators $\phi$, $\phi^{\dagger}$ and $N$ that satisfy the following commutation relations,
\small\begin{equation*}
[\phi,\phi^{\dagger}]= \pi \textrm{  ,  } [N,\phi]=-\phi \textrm{  ,  } [N,\phi^{\dagger}]=\phi^{\dagger}
\end{equation*}\normalsize
where $\pi = |0\rangle \langle0|$ is the vacuum projector. The one dimensional Fock space, $\mathbb{F}$, for this algebra is formed from the state $|n\rangle$, where the label $n \in \mathbb{Z}_+ \cup \{ 0\}$ is called an occupation number. The action of the operators on elements of the Fock space are given by,
\small\begin{equation*}
\phi^{\dagger}|n\rangle =|n+1\rangle \textrm{  ,  }\phi |n\rangle = |n-1\rangle \textrm{  ,  } N |n\rangle = n|n\rangle
\end{equation*}\normalsize
The action of the $\phi$ operator on the vacuum state, $|0\rangle$, annihilates it. \\
\indent We now extend the above bosonic algebra and consider the tensor product,
\small\begin{equation*}
\mathbb{F} = \mathbb{F}_0 \otimes \mathbb{F}_1 \otimes \dots \otimes \mathbb{F}_M
\end{equation*}\normalsize
We introduce the operators, $\phi_j$, $\phi^{\dagger}_j$ and $N_j$, $0 \le j \le M$, that act on $\mathbb{F}_j$ as,
\small\begin{equation*}
\phi_j = I_0 \otimes I_1 \otimes \dots I_{j-1} \otimes \phi \otimes I_{j+1} \otimes \dots \otimes I_M 
\end{equation*}\normalsize
and similarly for $\phi^{\dagger}_j$ and $N_j$, where $I_k$ is the identity operator in $\mathbb{F}_k$. The commutation relations are given by
\small\begin{equation}
[\phi_j,\phi^{\dagger}_k]= \pi_j \delta_{jk} \textrm{  ,  } [N_j,\phi_k]=-\phi_j \delta_{jk} \textrm{  ,  } [N_j,\phi^{\dagger}_k]=\phi^{\dagger}_j \delta_{jk}
\end{equation}\normalsize
where each operator of index $j$ acts on the corresponding indexed Fock vectors,
\small\begin{equation}\begin{array}{lcl}
\left(\phi_j \right)^{m_j-n_j} |m_j \rangle_j &=& |n_j \rangle_j \textrm{  for  }0 \le n_j < m_j \\
\left(\phi^{\dagger}_j \right)^{n_j-m_j} |m_j \rangle_j &=& |n_j \rangle_j \textrm{  for  } n_j > m_j \ge 0\\
N_j |m_j \rangle_j &=& m_j |m_j \rangle_j
\end{array}\end{equation}\normalsize
and $\phi_j$ annihilates the vacuum state $|0\rangle_j$. The state vectors, $|n_p \rangle_j$, and the corresponding conjugate vectors, $ \langle n_r|_k$, are orthonormal,
\small\begin{equation}
\langle n_r|n_p\rangle_{k,j}= \delta_{pr}\delta_{jk}
\end{equation}\normalsize
\subsection{Algebraic Bethe ansatz}
\noindent We define the phase model through the following local $L$-operator matrix,
\small\begin{equation}
L_j (u) \equiv \left( 
\begin{array}{cc}
\hat{a}_j (u) & \hat{b}_j (u)\\
\hat{c}_j (u) & \hat{d}_j (u)
\end{array}
 \right)= \left( 
\begin{array}{cc}
\frac{1}{u} & \phi^{\dagger}_j \\
\phi_j & u 
\end{array}
 \right)
\end{equation}\normalsize
where $u \in \mathbb{C}$. Naturally associated with $L_j(u)$ is the $4 \times 4$ $R$-matrix,
\small\begin{equation}
R(u,v) = \left( 
\begin{array}{cccc}
f(u,v) &0 & 0 & 0\\
0& g(u,v) & 1 & 0 \\
0 & 0 & g(u,v) & 0\\
0 & 0 & 0 & f(u,v)
\end{array}
 \right)
\end{equation}\normalsize
where $f (u,v) = \frac{u^2}{u^2-v^2}$ and $g (u,v) = \frac{u v}{u^2-v^2}$. $L$ and $R$ satisfy the following intertwining relation,
\small\begin{equation}
R(u,v)[L_j(u) \otimes L_j(v)] = [L_j(v) \otimes L_j(u)]R(u,v)
\label{inte}\end{equation}\normalsize\\
\textbf{The monodromy matrix.} The monodromy matrix, $T(u)$, for the phase model is introduced as the ordered product of all $(M+1)$ $L$-matrices,
\small\begin{equation}
T(u) = L_M(u) L_{M-1}(u) \dots L_0(u) = \left( 
\begin{array}{cc}
A (u) & B (u)\\
C (u) & D (u)
\end{array}
 \right)
\end{equation}\normalsize
Using induction on the intertwining relation (eq. \ref{inte}) the monodromy matrix and the $R$-matrix satisfy an equivalent intertwining relationship,
\small\begin{equation}
R(u,v)[T(u)\otimes T(v)] = [T(v) \otimes T(u)] R(u,v)
\label{inter.}\end{equation}\normalsize
which generate sixteen non trivial algebraic relationships. \\
\\
\textbf{Non local creation and annihilation operators.} Focusing on the operators $B(u)$ and $C(u)$, we consider the operator $\hat{N} = \sum^M_{j=0}N_j$, which measures the total occupation number of the state. Applying this operator to $B(u)$ and $C(u)$ we obtain,
\small\begin{equation}
\hat{N} B(u) = B(u) \left\{ \hat{N}+1\right\} \textrm{  ,  } \hat{N} C(u) = C(u) \left\{ \hat{N}-1 \right\}
\end{equation}\normalsize
Thus the operator $B(u)$ is a creation operator of the phase model, where one application on a state vector increases the total occupation number by one, while $C(u)$ is the opposing annihilation operator of the phase model, where one application to a state vector decreases the total occupation number by one. We note that $C(u)$ annihilates the total vacuum operator.\\
\indent Equivalently, the roles of $B(u)$ and $C(u)$ are reversed when applied to the conjugated vacuum vectors.  
\subsection{N-particle state vector and its conjugate.} We construct the $N$-particle vector, $|\Psi_M\rangle$, by repeated application of the construction operator $B$ on the vacuum vector,
\small\begin{equation*}
|\Psi_M \rangle = B(u_1) \dots B(u_N)|0\rangle
\end{equation*}\normalsize
where the total occupation number of $|\Psi_M \rangle$ is $N$. An alternative form for the $N$-particle vector is given by,
\small\begin{equation}
|\Psi_M \rangle = \sum_{\genfrac{}{}{0mm}{}{0 \le n_0,n_1,\dots,n_M \le N}{n_0 + n_1 +\dots + n_M = N}}f_{\{n_0,\dots,n_M \}}(\vec{u}) \prod^M_{k=0} \left(\phi^{\dagger}_k \right)^{n_k} \bigotimes^{M}_{j=0} |0\rangle_j = \sum_{\{\lambda \} \subseteq (M)^N}f_{\{ \lambda \}}(\vec{u}) | \lambda \rangle  
\label{statevec}\end{equation}\normalsize
where the partition $\{\lambda \}$ is constructed from the occupation number sequence in the following manner,
\small\begin{equation}
\{\lambda\} = (M^{n_M}, (M-1)^{n_{M-1}}, \dots, 1^{n_1},0^{n_0})
\label{part.}\end{equation}\normalsize
Similarly, we construct the conjugate $N$-particle vector, $\langle \Psi_M |$, by repeated application of the annihilation operator $C$ on the conjugate vacuum vector,
\small\begin{equation*}\begin{split}
\langle \Psi_M | = \langle 0| C(v_N) \dots C(v_1)
\end{split}\end{equation*}\normalsize
where the total occupation number $N$, and an alternative form is given by,
\small\begin{equation}
\langle \Psi_M |= \sum_{\genfrac{}{}{0mm}{}{0 \le n_0,n_1,\dots,n_M \le N}{n_0 + n_1 +\dots + n_M = N}}g_{\{n_0,\dots,n_M \}}(\vec{v})  \bigotimes^{M}_{j=0} \langle 0|_j \prod^M_{k=0} \left(\phi_k \right)^{n_k}= \sum_{\{\lambda \} \subseteq (M)^N}g_{\{ \lambda \}}(\vec{v}) \langle \lambda |
\label{conjstatevec}\end{equation}\normalsize
\textbf{Schur polynomial expansion of state vectors.} In \cite{Tsilevich}, Tsilevich derived the following Schur polynomial forms for $f_{\{ \lambda \}}(\vec{u})$ and $g_{\{ \lambda \}}(\vec{v})$ in eqs. \ref{statevec} and \ref{conjstatevec},
\small\begin{equation}
f_{\{ \lambda \}}(\vec{u}) = \left(\prod^N_{j=1} u_j \right)^{-M} S_{\{ \lambda \}} (u^2_1,\dots,u^2_N) \textrm{  ,  }g_{\{ \lambda \}}(\vec{v}) = \left(\prod^N_{j=1} v_j \right)^M S_{\{ \lambda \}} (v^{-2}_1 ,\dots,v^{-2}_N ) 
\label{Tsil}\end{equation}\normalsize
In order to proceed, we need to express the state vectors as weighted sums of lattice configurations, plane partitions and semi-standard tableaux. 
\subsection{Weighted sums of lattice paths.} 
\begin{figure}[h!]
\begin{center}
\includegraphics[angle=0,height=55mm,width=90mm]{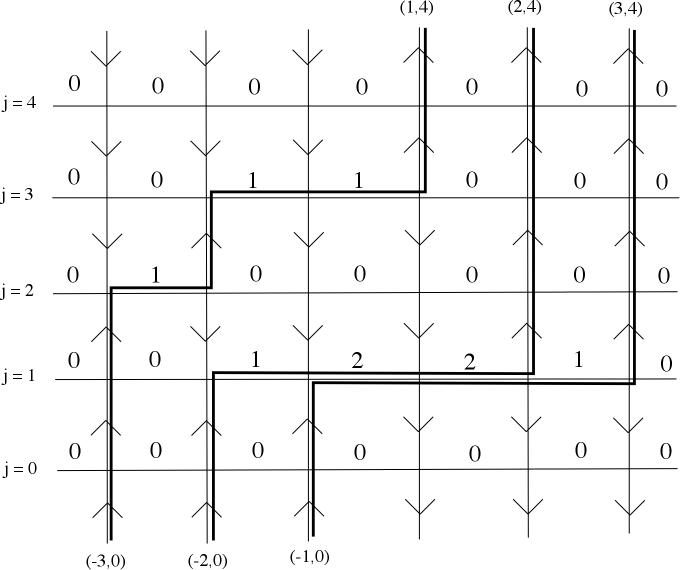}
\caption{\footnotesize{Generic lattice path configuration for $M =4$, $N = 3$.}}
\label{1.c}
\end{center}
\end{figure}

\begin{figure}[h!]
\begin{center}
\includegraphics[angle=0,scale=0.30]{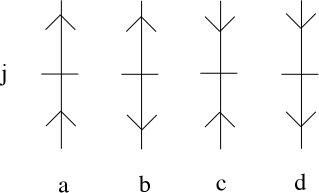}
\caption{\footnotesize{The four vertices used to construct lattice paths.}}
\label{1.a}
\end{center}
\end{figure}

Consider $N$ non-crossing column strict lattice paths on the $(M+1)\times 2N$ lattice, where the paths begin at the first $N$ bottom-most horizontal edges, $(-N,0), (-N+1,0), \dots, (-1,0)$, and end at the final $N$ top-most horizontal edges, $(1,M), (2,M), \dots, (N,M)$, respectively. Each path is restricted to move either up or right, and no two paths can cross or occupy the same vertical edge.  We give a typical example in fig. \ref{1.c}.\\
\\
Given an allowable lattice path configuration, we assign each of the four possible vertices a letter, as indicated in fig. \ref{1.a}.\\
\\
Following section V of \cite{Bog2}, it is possible to express the coefficient of the conjugate state vector, $g_{\{n_{0},\dots,n_{M} \}}(\vec{v})$, as the following weighted sum of lattice paths, 
\small\begin{equation}
g_{\{n_{j_1},\dots,n_{j_k} \}}(\vec{v}) = \sum_{\genfrac{}{}{0mm}{}{\textrm{allowable paths}}
{\textrm{in $(M+1)\times N$ lattice}}} v_1^{t^d_1-t^a_1} v_2^{t^d_2-t^a_2} \dots v_N^{t^d_N-t^a_N}
\end{equation}\normalsize
where the sum is taken over all allowable paths in the $(M+1) \times N$ lattice under the conditions,
\begin{itemize}
\item{$n_{j_1}$ paths start at $(-N,0),(-N-1,0), \dots, (-N+n_{j_1}-1,0)$ and end at $(-1,j_1)$.}
\item{$n_{j_2}$ paths start at $(-N+n_{j_1} ,0),\dots,(-N+\sum^2_{l=1}n_{j_l}-1,0)$ and end at $(-1,j_2)$.}
\item{This procedure continues until we finally have $n_{j_k}$ paths starting at $(-N+\sum^{k-1}_{l=1}n_{j_l} ,0),\dots,(-1,0)$ and ending at $(-1,j_k)$.}
\end{itemize}
The powers $t^d_{l}$ and $t^a_{l}$, $1 \le l \le N$, are equal to the number of $d$ and $a$ vertices respectively in the $l$th column.\\
\indent Similarly, the coefficient of the state vector, $f_{\{n_{0},\dots,n_{M} \}}(\vec{u})$, can be expressed as the following weighted sum of lattice paths, 
\small\begin{equation}
f_{\{n_{j_1},\dots,n_{j_k} \}}(\vec{u}) = \sum_{\genfrac{}{}{0mm}{}{\textrm{allowable paths}}{\textrm{in $(M+1)\times N$ lattice}}}u_1^{t^d_1-t^a_1} u_2^{t^d_2-t^a_2} \dots u_N^{t^d_N-t^a_N}
\end{equation}\normalsize
where the sum is taken over all allowable paths in the $(M+1) \times N$ lattice under the conditions,
\begin{itemize}
\item{$n_{j_1}$ paths start at $(1,j_1)$ and end at $(1,M),(2,M),\dots,(n_{j_1},M).$}
\item{$n_{j_2}$ paths start at $(1,j_2)$ and end at $(n_{j_1} +1,M),\dots,(\sum^2_{l=1}n_{j_l},M).$}
\item{This procedure continues until we finally have $n_{j_k}$ paths starting at $(1,j_k)$ and ending at $(\sum^{k-1}_{l=1}n_{j_l} +1,M),\dots,(N,M)$.}
\end{itemize}
\subsection{Weighted sums of plane partitions.} 
\begin{figure}[h!]
\begin{center}
\includegraphics[angle=0,height=42mm,width=42mm]{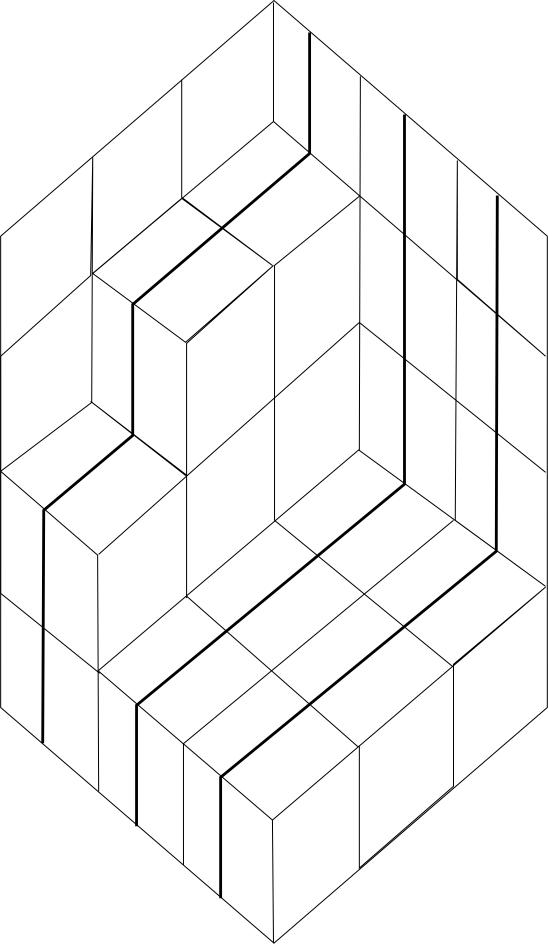}
\caption{\footnotesize{Graphical representation of the plane partition $\pi^{\{\lambda'\}}$.}}
\label{1.d}
\end{center}
\end{figure}

\begin{figure}[h!]
\begin{center}
\includegraphics[angle=0,scale=0.13]{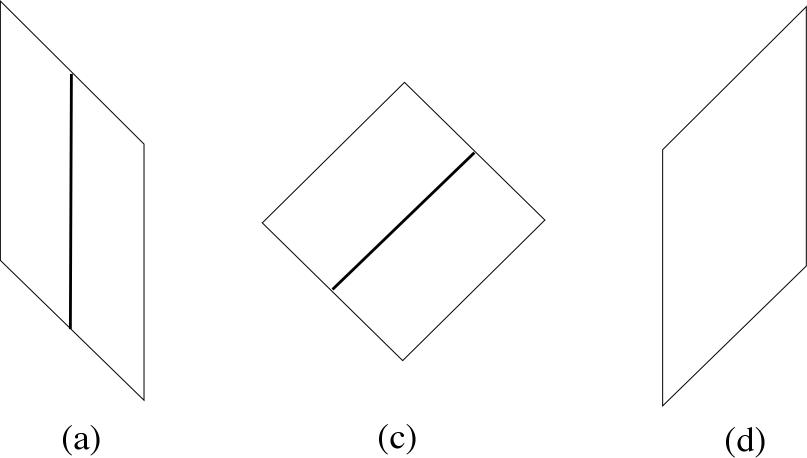}
\caption{\footnotesize{The three types of rhombi used to construct plane partitions.}}
\label{1.e}
\end{center}
\end{figure}

A plane partition, $\pi_{j,k}$, is an array of non negative integers such that,
\small\begin{equation*}
\pi_{j,k} \ge \pi_{j+1,k} \textrm{ and  } \pi_{j,k} \ge \pi_{j,k+1}
\end{equation*}\normalsize
If we restrict the size of the array to be $N \times N$, and restrict the maximum of any integer within the plane partition, $\pi_{i,j} \le M$, the plane partition is said to be contained within a box of side lengths $N \times N \times M$.\\
\indent A typical example of a plane partition within a box of $3 \times 3 \times 4$ is given by the following\footnote{We shall use the following plane partition, $\pi^{\{\lambda'\}}$, as a running example in this section.},
\small\begin{equation}
\pi^{\{\lambda'\}}=\left( \begin{array}{ccc}
3 & 1 & 1\\
3 & 1 & 1\\
2 & 1 & 1  
\end{array}\right)
\label{ppexa}\end{equation}\normalsize
Note that the diagonal entries of any plane partition always give a partition in usual sense\footnote{In the above example we obviously have $\{\lambda'\} = (3,1,1).$}. The graphical representation of a plane partition in a $N \times N \times M$ box is given by considering rhombus tilings of a $(N, N, M)$ semiregular hexagon. The plane partition, $\pi^{\{\lambda'\}}$, is represented by fig. \ref{1.d}, where each representation can be generally constructed entirely from the three types of rhombi given in fig. \ref{1.e}.\\
\newpage
\noindent \textbf{Correspondence between plane partitions in a $\mathbf{N \times N \times M}$ box and $\mathbf{N}$ non-crossing column strict lattice paths on the $\mathbf{(M+1)\times 2N}$ lattice.} The $j$th path of the lattice configuration can be thought of as the $j$th column of the array $\pi^{\lambda}$. As an example, consider the array, $\pi^{\{\lambda'\}}$, which is in correspondence with the lattice path configuration shown in fig. \ref{1.c}.\\
\indent The bottom left entry, $\pi^{\{\lambda'\}}_{3,1}=2$, corresponds to the first horizontal section of the first path on the second row, and the remaining entries of the column, $(\pi^{\{\lambda'\}}_{2,1}=3,\pi^{\{\lambda'\}}_{1,1}=3)$, correspond to the remaining horizontal sections of the first path, on the third row.\\
\indent There exists a similar correspondence between the second and third lattice paths, and the second and third columns of the array $\pi^{\{\lambda'\}}$ respectively. It should be clear how this process is generalized for any $N$ and $M$.\\
\indent Due to the above correspondence, when considering \textit{lower diagonal plane partitions in an $N\times N \times M$ box}, (equivalently, the \textit{left hand side of the $(N,N,M)$ rhombus tiling}), we obtain
\small\begin{equation}
g_{\{n_{j_1},\dots,n_{j_k} \}}(\vec{v}) = \sum_{\genfrac{}{}{0mm}{}{\textrm{lower diagonal} }{\textrm{plane partitions}}}v_1^{l^d_1-l^a_1} v_2^{l^d_2-l^a_2} \dots v_N^{l^d_N-l^a_N}
\end{equation}\normalsize
where the sum is taken over all allowable lower diagonal $N \times N\times M$ plane partitions, (left hand side $(N,N,M)$ rhombus tilings), and the diagonal terms are given by the partition representation of the corresponding occupation number sequence (eq. \ref{part.}). The powers $l^d_{l}$ and $l^a_{l}$, $1 \le l \le N$, are equal to the number of $d$ and $a$ rhombi respectively in the $l$th column of the left half rhombus tiling.\\
\indent Similarly, when considering \textit{upper diagonal plane partitions in an $N\times N \times M$ box}, (equivalently, the \textit{right hand side of the $(N,N,M)$ rhombus tiling}), we obtain
\small\begin{equation}
f_{\{n_{j_1},\dots,n_{j_k} \}}(\vec{u}) = \sum_{\genfrac{}{}{0mm}{}{\textrm{upper diagonal}}{\textrm{plane partitions}}} u_1^{l^d_1-l^a_1} u_2^{l^d_2-l^a_2} \dots u_N^{l^d_N-l^a_N}
\end{equation}\normalsize
where the sum is taken over all allowable upper diagonal $N\times N \times M$ plane partitions. Again, the diagonal terms are given by the partition representation of the corresponding occupation number sequence. 
\subsection{Weighted sums of semi-standard tableaux.} 
\begin{figure}[h!]
\begin{center}
\includegraphics[angle=0,scale=0.20]{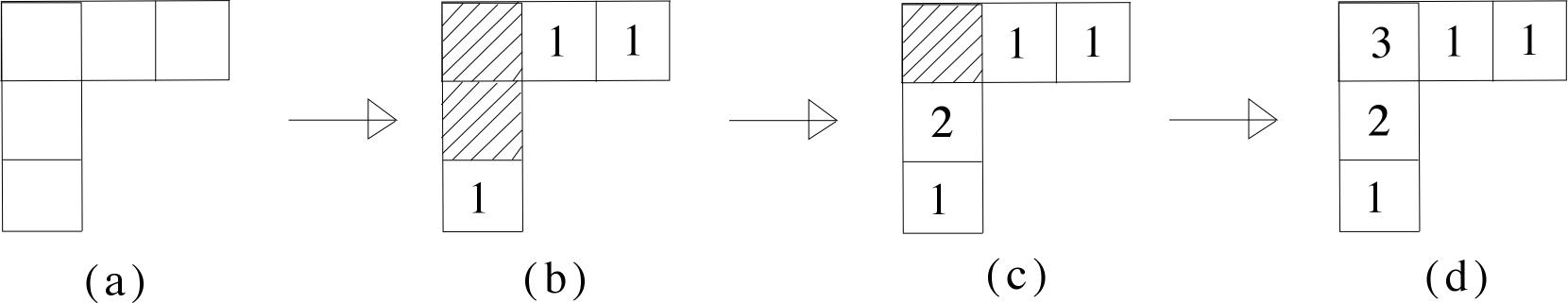}
\caption{\footnotesize{Construction of the tableau $T^{\{\lambda' \}}_-$.}}
\label{1.f}
\end{center}
\end{figure}

\begin{figure}[h!]
\begin{center}
\includegraphics[angle=0,scale=0.20]{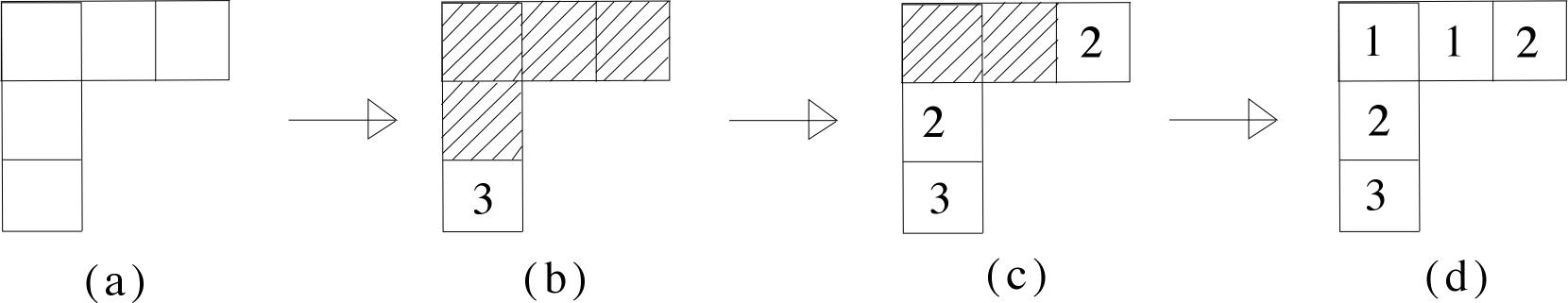}
\caption{\footnotesize{Construction of the tableau $T^{\{\lambda' \}}_+$.}}
\label{1.g}
\end{center}
\end{figure}

We now give the final alternative forms of the state vectors using the following correspondences.\\
\\
\textbf{Correspondence between upper diagonal plane partitions, $\mathbf{\pi^{\{ \lambda\}}_+}$, and semi-standard tableaux of descending order\footnote{Semi-standard tableaux are commonly of ascending numerical order, however, descending numerical order is the most convenient convention for the purposes of the next section.}, $\mathbf{T^{\{\lambda \}}_-}$.} We begin by considering a general upper half plane partition, $\pi^{\{\lambda \}}_{+}$, and construct a partition using the diagonal entries,
\small\begin{equation*}
\{ \lambda \} = \{\pi_{1,1},\pi_{2,2},\dots, \pi_{N,N}\}
\end{equation*}\normalsize
Considering the next upper diagonal entries, $\pi_{j,j+1}$, we construct the skew diagram, $\{ \mu_1 \}$,
\small\begin{equation*}
\{ \mu_1 \}= \{\pi_{1,1}-\pi_{1,2},\pi_{2,2}-\pi_{2,3},\dots, \pi_{N-1,N-1}-\pi_{N-1,N},\pi_{N,N}\}
\end{equation*}\normalsize
and place the integer $1$ in the valid regions of the skew diagram\footnote{In ascending tableaux, $N$ would be placed instead of $1$.}. We then consider the next upper diagonal entries of the array, $\pi_{j,j+2}$, and construct the skew diagram, $\{ \mu_2 \}$,
\small\begin{equation*}
\{ \mu_2 \} = \{\pi_{1,1}-\pi_{1,3},\pi_{2,2}-\pi_{2,4},\dots, \pi_{N-2,N-2}-\pi_{N,N-2},\pi_{N-1,N-1},\pi_{N,N}\}
\end{equation*}\normalsize
and place the integer $2$ in the valid regions of the skew diagram that have not already been occupied by previous steps in this process. This process continues until the partition contains the numbers $\{1,\dots,N-1 \}$. We then fill the remaining boxes in the partition with the integer $N$, thereby constructing a valid descending semi-standard tableau $T^{\{\lambda \}}_-$ from the upper diagonal plane partition $\pi^{\{\lambda \}}_{+}$.\\
\indent As an example, consider the array, $\pi^{\lambda'}$, given in the past examples where $\{\lambda' \} = (3,1,1)$. The construction of the corresponding descending semi-standard tableau is given in fig. \ref{1.f}. \\
\indent In \textbf{(a)} we construct the partition $\{\lambda'\} = (3,1,1)$. In \textbf{(b)} we construct the skew partition $\{ \mu_1\}=(3,1,1)-(1,1,0)$ and place the integer 1 in the valid regions of $\{ \mu_1\}$. The partition $(1,1,0)$ was obtained from the first upper diagonal entries of $\pi^{\lambda'}$. In \textbf{(c)} we construct the skew partition $\{ \mu_2\}=(3,1,1)-(1,0,0)$ and place the integer 2 in the valid regions of $\{ \mu_2\}$ that contain no integers. The partition $(1,0,0)$ was obtained from the second upper diagonal entries of $\pi^{\lambda'}$. In $\textbf{(d)}$ we place the integer 3 in any remaining entries of $\{\lambda'\}$ that don't already contain integers, forming the valid descending semi-standard tableau $T^{\{\lambda' \}}_-$ from the upper diagonal plane partition $\pi^{\{\lambda' \}}_{+}$.\\
\\
Thus based on the above correspondence, another valid combinatorial definition for $ f_{\{\lambda \}}(\vec{u})$ is the following,
\small\begin{equation}
f_{\{\lambda \}}(\vec{u}) = \sum_{T^{\{\lambda \}}_-}u_1^{2t_1 - M} u_2^{2t_2 - M} \dots u_N^{2t_N - M}
\end{equation}\normalsize
where the summation is over all semi-standard Young tableaux of shape $\{\lambda\}$. The powers, $t_j$, give the weights of $T^{\{\lambda\}}_-$, which count the number of times $j$ appears in the tableau. Note that this expression is in accordance with eq. \ref{Tsil}.\\
\\
\textbf{Correspondence between lower diagonal plane partitions, $\mathbf{\pi^{\{ \lambda\}}_-}$, and semi-standard tableaux of ascending order, $\mathbf{T^{\{\lambda \}}_+}$.} Using an equivalent algorithm as described above, except this time applying a numerically ascending convention, we obtain the required correspondence. Using $\pi^{\{\lambda'\}}$ as an example yet again, the construction of the corresponding ascending semi-standard tableau is given in fig. \ref{1.g}.\\
\indent In \textbf{(a)} we construct the partition $\{\lambda'\} = (3,1,1)$. In \textbf{(b)} we construct the skew partition $\{ \nu_1\}=(3,1,1)-(3,1,0)$ and place the integer 3 in the valid regions of $\{ \nu_1\}$. The partition $(3,1,0)$ was obtained from the first lower diagonal entries of $\pi^{\lambda'}$. In \textbf{(c)} we construct the skew partition $\{ \nu_2\}=(3,1,1)-(2,0,0)$ and place the integer 2 in the valid regions of $\{ \nu_2\}$ that contain no integers. The partition $(2,0,0)$ was obtained from the second lower diagonal entries of $\pi^{\lambda'}$. In $\textbf{(d)}$ we place the integer 1 in any remaining entries of $\{\lambda'\}$ that don't already contain integers, forming the valid ascending semi-standard tableau $T^{\{\lambda' \}}_+$ from the lower diagonal plane partition $\pi^{\{\lambda' \}}_{-}$.\\
\\
Thus we immediately obtain,
\small\begin{equation}
g_{\{\lambda \}}(\vec{v}) = \sum_{T^{\{\lambda \}}_+}v_1^{-2t_1 + M} v_2^{-2t_2 + M} \dots v_N^{-2t_N + M}
\end{equation}\normalsize
where the summation is over all semi-standard Young tableaux of shape $\{\lambda\}$ of ascending numerical order. 
\subsection{The scalar product.} We now consider the scalar product, $\field{S}(N,M| \vec{u},\vec{v})$, of the phase model which is defined as the inner product of the state vectors,
\small\begin{equation}
\field{S}(N,M| \vec{u},\vec{v}) = \langle \Psi_M |\Psi_M \rangle
\label{scalarfirst'}
\end{equation}\normalsize
Using the algebraic expressions from eq. \ref{inter.}, it is possible to obtain the following determinant expression,
\small\begin{equation}
\field{S}(N,M) = \left( \prod^N_{i=1}\frac{1}{u_i v_i} \right)^{M} \prod_{1 \le j < k \le N} \left( \frac{1}{u^2_j-u^2_k}\right) \left( \frac{1}{v^2_j-v^2_k}\right)  \textrm{det} \left[h_{M+N-1}(u^2_m,v^2_l)\right]_{1 \le m,l \le N}
\label{scalardet}\end{equation}\normalsize
Alternatively, considering the Schur polynomial expansion of the state vectors,
\small\begin{equation}
\field{S}(N,M) = \left(\prod^N_{j=1} \frac{v_j}{u_j} \right)^{M}  \sum_{\{\lambda \} \subseteq (M)^N}S_{\{ \lambda \}} (u^2_1,\dots,u^2_N) S_{\{ \lambda \}} (v^{-2}_1 ,\dots,v^{-2}_N )
\label{scalarfirst}
\end{equation}\normalsize
\subsection{Restricting the 2-Toda tau-function to obtain the scalar product.}
\begin{proposition}
\label{illu}
The scalar product of the phase model for general $N$ and $M$ is, up to an overall factor of $ \left( \prod^N_{j=1} \frac{v_j}{u_j} \right)^M$, a restricted $\tau$-function of the 2-Toda hierarchy with $A_{\{\lambda\}\{\mu\}} = \delta_{\{\lambda\}\{\mu\}}$ and $s=n-M=m+N$, where $m$ and $n$ are free parameters.
\end{proposition}
\textbf{Proof.} Beginning with the unrestricted $\tau$-function,
\small\begin{equation*}
\tau(s = n-M=m+N, \vec{x},\vec{y}) = \sum_{\{\lambda\} \subseteq (M)^N} \chi_{\{\lambda\} }(\vec{x}) \chi_{\{\lambda\} } (-\vec{y})
\end{equation*}\normalsize
and performing the following change of variables,
\small\begin{equation}
x_k \rightarrow \frac{1}{k} p_k\left(u^2_1, \dots,u^2_N  \right)\textrm{  ,  } -y_k \rightarrow \frac{1}{k} p_k\left(v^{-2}_1, \dots,v^{-2}_N  \right)\textrm{  ,  } 1 \le k \le N+M-1 
\label{usual}\end{equation}\normalsize
we obtain the required result. $\square$\\
\indent The above result only considers one value of $s$. Let us now consider the family of corresponding restricted $\tau$-functions for other values of $s$. We begin by clarifying some known facts about the family of unrestricted $\tau$-functions. 
\begin{itemize}
\item{The entire family is given by $\{\tau_{s=m+1}(\vec{x},\vec{y}), \tau_{s=m+2}(\vec{x},\vec{y}), \dots, \tau_{s=n}(\vec{x},\vec{y})\}$.}
\item{Different values of  $s$ \textbf{do not} change the amount of, (two sets of $n-m-1$), time variables.}
\end{itemize}
We now compare this to the case of the family of restricted $\tau$-functions. 
\begin{itemize}
\item{The initial $\tau$-function, $\tau\left(s = n-M=m+N, \left\{ u^2 \right\}, \left\{ v^{-2} \right\} \right)$, has two sets of $N+M-1$ time variables, but each set is constructed from $N$ symmetric variables.}
\item{The introduction of the  condition $s = n-M=m+N$ means that as $s$ changes, so to do $M$ and $N$. By considering the change in the dimensions of the partition, we can obtain how $M$ and $N$ change with $s$.}
\end{itemize}
\small\begin{equation}
s \rightarrow s \pm l \Longleftrightarrow  \left\{ \begin{array}{c} M \rightarrow M \mp l \\ N \rightarrow N \pm l \end{array}\right. 
\label{changeins}\end{equation}\normalsize
\begin{itemize}
\item{Consequently, although the number of time variables do not change with each $s$ value, different values of  $s$ \textbf{do} change the amount of symmetric variables that the time variables are constructed from.}
\end{itemize}
\textbf{An example.} Consider the complete family of unrestricted $\tau$-functions for $n=5$ and $m=1$. In this case each $\tau$-function contains two sets of $3$ time variables, $\{\vec{x},\vec{y} \}=\{x_1,x_2,x_3,y_1,y_2,y_3\}$,
\small\begin{equation*}
\left\{ \tau_{s=2}  = \sum_{\{\lambda\} \subseteq \{3 \}}\Theta_{\{\lambda \}} \textrm{  ,  }  \tau_{s=3} = \sum_{\{\lambda\} \subseteq \{2,2 \}}\Theta_{\{\lambda \}} \textrm{  ,  } \tau_{s=4}  = \sum_{\{\lambda\} \subseteq \{1,1,1 \}} \Theta_{\{\lambda \}}  \textrm{  ,  } \tau_{s=5} =\Theta_{\{\phi \}} \right\}
\end{equation*}\normalsize
where $\Theta_{\{\lambda \}}= \chi_{\{\lambda\}}(\vec{x})\chi_{\{\lambda\}}(-\vec{y})$. The main question now is, if one $\tau$-function in a family has been restricted to form a scalar product with a certain $M$ and $N$ value, \emph{can the remaining $\tau$-functions of the family also be restricted to form scalar products with valid $M$ and $N$ values?}\\
\indent Naively performing the corresponding restrictions to the above family of $\tau$-functions we obtain the following family of scalar products,
\small\begin{equation*}
\left\{ \gamma^{3}_{1} \field{S}\left(\genfrac{}{}{0mm}{}{N=1}{M=3} \right), \gamma^{2}_{2} \field{S}\left(\genfrac{}{}{0mm}{}{N=2}{M=2}\right), \gamma^{1}_{3}  \field{S}\left(\genfrac{}{}{0mm}{}{N=3}{M=1} \right), \gamma^{0}_{4} \field{S}\left(\genfrac{}{}{0mm}{}{N=4}{M=0}  \right) \right\}
\end{equation*}\normalsize
where $\gamma^M_N =\left( \prod^N_{j=1} \frac{u_j}{v_j} \right)^M$. This example illustrates an important issue. We remember that each $\tau$-function contained within a family must contain the same amount of time variables. Furthermore, it is a requirement that these time variables be the same for each value of $s$ to ensure that the $\tau$-functions obey the bilinear relation\footnote{The $\tau$-functions obviously \textbf{must} obey the bilinear relation.}. If this is to be the case for the above example, we have the following set of equations that must be satisfied,
\small\begin{equation*}\begin{split}
 p_k\left((u^i_1)^2  \right)= p_k\left((u^{ii}_1)^2, (u^{ii}_2)^2  \right)=  \dots= p_k\left((u^{iv}_1)^2, \dots,(u^{iv}_4)^2  \right)\\
p_k\left((v^i_1)^{-2}  \right) = p_k\left((v^{ii}_1)^{-2},(v^{ii}_2)^{-2}  \right) =  \dots = p_k\left((v^{iv}_1)^{-2},\dots,(v^{iv}_4)^{-2} \right)
\end{split}\end{equation*}\normalsize
for $1 \le k \le 3$. A simple check will reveal that only the trivial solution exists, meaning that all but one of the symmetric variables are set to zero. Thus, at a first glance, the answer to the question is no, due to the fact that \emph{the $\tau$-functions in the family all need to contain the same time variables, be they independent or restricted}.\\
\indent We now generalize the above example.
\begin{proposition}
\label{ref1}
The system of equations, $0 \le l \le M-1$,
\small\begin{equation*}\begin{split}
u_1^2+ \dots +u_{N+l}^{2} & =\mu_1^2+ \dots +\mu_{N}^{2}\\
u_1^4+ \dots +u_{N+l}^{4} & =\mu_1^4+ \dots +\mu_{N}^{4} \\
&\vdots\\
u_1^{2(N+M-1)}+ \dots +u_{N+l}^{2(N+M-1)} & = \mu_1^{2(N+M-1)}+ \dots +\mu_{N}^{2(N+M-1)}
\end{split}\end{equation*}\normalsize
permits only the trivial solution, i.e. $u_{\sigma_j}^2= \mu_j^2$, for $j\in \{1,\dots,N\}$, and the remaining $l$ of the $u_k^2$'s are equal to zero. 
\end{proposition}
\textbf{Proof.} We begin by considering the first $N+l$ equations in the system, the remaining equations will follow easily. We note that the left hand side of these polynomial equations exist in the symmetric polynomial ring $\mathbb{C}[u_1^2,\dots,u_{N+l}^2]^{S_{N+l}}$. Consider now the polynomial ring $\mathbb{C}[s_1,\dots,s_{N+l}]$, and recall that the fundamental theorem of symmetric polynomials states that there exists an isomorphism between the two rings, $\mathbb{C}[u_1^2,\dots,u_{N+l}^2]^{S_{N+l}} \cong \mathbb{C}[s_1,\dots,s_{N+l}]$, with the isomorphism sending\footnote{We could use any basis symmetric polynomial, $e_j,h_j,p_j$, for the mapping. See section I.2 of \cite{MacD} for further details.} $p_j\left(u_1^2,\dots,u_{N+l}^2\right) \rightarrow s_j$, $j=\{1,\dots,N+l\}$. Hence the system of $N+l$ equations in the isomorphic polynomial ring, $\mathbb{C}[s_1,\dots,s_{N+l}]$, is linear and has one solution.\\
\indent Thus in the ring $\mathbb{C}[u_1^2,\dots,u_{N+l}^2]^{S_{N+l}}$, the system contains one base solution, and every possible permutation of that base solution (since the polynomial ring is symmetric), leading to a total of $(N+l)!$ possible solutions. Since we already trivially know $(N+l)!$ solutions to the system, $u_{\sigma_j}^2= \mu_j^2$ for $j\in \{1,\dots,N\}$, and $u_{\sigma_k}^2= 0$ for $k\in \{N+1,\dots,N+l\}$, this means only the trivial solution exists for the first $N+l$ equations. It remains to note that the remaining $M-l-1$ equations are solved by the $(N+l)!$ solutions. $\square$\\
\indent Using the above result the following lemma comes almost automatically.
\begin{lemma}
Assume we have a particular family of unrestricted $\tau$-functions with particular $m$ and $n$ values,
\small\begin{equation}
\{\tau_{m+1}(\vec{x},\vec{y}), \tau_{m+2}(\vec{x},\vec{y}), \dots, \tau_{n}(\vec{x},\vec{y})\}
\label{taufamily}\end{equation}\normalsize
The process of restricting the entire family so that each $\tau$-function corresponds to a valid scalar product expression,
\small\begin{equation}
\{\gamma^{n-m-1}_1  \field{S}\left(\left.\genfrac{}{}{0mm}{}{N=1}{M=n-m-1}\right| \vec{u},\vec{v} \right), \dots, \gamma^{0}_{n-m}  \field{S}\left(\left.\genfrac{}{}{0mm}{}{N=n-m}{M=0}\right| \vec{\mu},\vec{\nu} \right)\}
\label{scafamily}\end{equation}\normalsize
has potentially two (ill) effects.
\begin{itemize}
\item{If each of the above scalar product expressions has two sets of $N$ ($N$ is not constant for each scalar product) symmetric variables, then the 2 sets of $n-m-1=N+M-1$ time variables of the restricted $\tau$-functions are no longer equal, and therefore the \textbf{bilinear identity is no longer valid}.}
\item{If we enforce that the time variables be equal, then we only have two sets of \textbf{one symmetric variable} for each of the scalar product expressions.}
\end{itemize} 
Arguably both scenarios are pointless, so it makes sense to use the results of proposition \ref{illu} and only consider restricting one $\tau$-function in any family.
\end{lemma}
\textbf{Proof.} Applying proposition \ref{illu} on all the unrestricted $\tau$-functions in eq. \ref{taufamily}, we instantly arrive to the expression in eq. \ref{scafamily}. Analyzing any two of the above scalar product expressions, (with two sets of $N$ symmetric variables), the results of proposition \ref{ref1} state that the symmetric power sums, and hence the time variables, cannot be equal. Thus the first point in this lemma becomes obvious. Furthermore, from proposition \ref{ref1}, the only way for the time variables to be equal is if we trivialize the power sums as indicated in point 2 of this lemma. $\square$
\section{The Toda wave-vectors}
In this section we shall show that restricting the wave-functions of 2-Toda hierarchy give an alternative method to calculating certain classes of correlation functions, and thus have a natural combinatorial meaning. In order to proceed we shall first give necessary definitions of skew Schur polynomials\footnote{For more details see section I.V of \cite{MacD}.}.\\
\\
\textbf{Skew Schur polynomials.} Given a set of variables $\vec{u} = (u_1,\dots,u_N)$ and the partitions $\{\lambda\}$, $\{\mu\}$, such that $\{\lambda\} \supseteq \{ \mu\}$, the skew Schur polynomial, $S_{\{\lambda\} / \{ \mu\}} (\vec{u})$, is defined as,
\small\begin{equation}\begin{split}
S_{\{\lambda\} /\{\mu\}}(\vec{u}) & = \sum_{T^{\{ \lambda - \mu \}}_+} u^{t_1}_1u^{t_2}_2 \dots u^{t_N}_N=\sum_{T^{\{ \lambda - \mu \}}_-} u^{t_1}_1u^{t_2}_2 \dots u^{t_N}_N\\
&= \textrm{det}[h_{\lambda_i - \mu_j +j-i}(u_1,\dots,u_N)]_{1 \le i,j \le N}
\end{split}\label{schurdef}\end{equation}\normalsize
where the sum is given over all possible (ascending or descending) semi-standard skew tableaux of shape $\{\lambda - \mu \}$, and the $t_j$ give the weights of the tableau.
\subsection{Wave-functions - I} Considering the $\hat{w}^{(0)}$ class of wave-functions, from eq. \ref{ned} we obtain,
\small\begin{equation*}
\tau(s) \hat{w}^{(0)}_{k}(s) = \sum_{\{\lambda \} \subseteq (n-(s+1))^{((s+1)-m)}} \chi_{\{\lambda \}} (\vec{x}) \chi_{\{\lambda \}/\{k \}} (-\vec{y}) 
\end{equation*}\normalsize
where we have used the following result,
\small\begin{equation}
\zeta_j (-\tilde{\partial}_{\vec{y}}) \chi_{\{\lambda \}}(-\vec{y}) = \chi_{\{\lambda \}/\{j\}} (-\vec{y})
\label{nice1}\end{equation}\normalsize
for all partitions $\{\lambda\}$ such that $\{j \} \subseteq \{\lambda\}$.\\
\indent Thus the upper triangular wave-matrix, $\hat{W}^{(0)}$, has entries of the form,
\small\begin{equation*}
\hat{W}^{(0)}= \left( \frac{1}{\tau(j)}\sum_{\{\lambda \} \subseteq (n-(j+1))^{((j+1)-m)}} \chi_{\{\lambda \}} (\vec{x}) \chi_{\{\lambda \}/\{k-j \}} (-\vec{y}) \right)_{m \le j,k \le n-1}
\end{equation*}\normalsize\\
\textbf{Constructing skew $N$-particle conjugate state vectors.} Consider the following conjugate state vector,
\small\begin{equation*}
\langle 0| \phi_k C(v_2)  \dots C(v_{N}) = \langle \Psi^{\{k \}}_M|
\end{equation*}\normalsize
The allowable partitions of this conjugate vector are given by the following result,
\begin{proposition}
\small\begin{equation*}
\langle \Psi^{\{k \}}_M |= \sum_{\genfrac{}{}{0mm}{}{\{\lambda\} \subseteq \{(M)^{ (N-1)},k\}}{\{\lambda\} \supseteq  \{ k \}}} \psi^{(1,k)}_{\{\lambda \}} \langle \lambda|
\end{equation*}\normalsize
\end{proposition}
\textbf{Proof.} Consider the non crossing column strict lattice path interpretation of the state vectors. The operator $\phi_k$ assures us that the first path in the first column makes a directional change from up to right at row $k$. This has the effect that the occupation number sequence will contain at least one entry $n_l$, where $l \ge k$. Transforming the occupation number sequence to a partition $\{ \lambda \}$, we instantly receive the result, $\{ \lambda \} \supseteq \{ k \}$.\\
\indent The fact that the first path in the first column turns right at row $k$ also means that the highest row that the $N$th path can be when it crosses between column $N$ and $N+1$ is $k$. Thus the highest partition obtainable from lattice paths under this restriction are $\{\lambda \} =  \{(M)^{(N-1)},k\}.$   $\square$\\
\\
\textbf{Combinatorial definitions of $\mathbf{\psi^{(1,k)}_{\{\lambda \}}}$.} Considering the \textbf{lattice path} interpretation we receive,
\small\begin{equation*}
\psi^{(1,k)}_{\{\lambda \}} = \sum_{\genfrac{}{}{0mm}{}{\textrm{allowable paths in}}{\textrm{$(M+1)\times N$ lattice$^\dagger$}}} v^{t^{d}_2-t^{a}_2}_2 \dots v^{t^{d}_N-t^{a}_N}_N
\end{equation*}\normalsize
where the lattice paths are under the condition that the first path in the first column makes a directional change from up to right at row $k$.\\
\indent Considering the \textbf{lower diagonal plane partition} interpretation we receive,
\small\begin{equation*}
\psi^{(1,k)}_{\{\lambda \}} = \sum_{\genfrac{}{}{0mm}{}{\textrm{lower diag. plane part.}}{\textrm{in $N \times N\times M$ array$^\dagger$}}}  v^{l^{d}_2-l^{a}_2}_2 \dots v^{l^{d}_N-l^{a}_N}_N
\end{equation*}\normalsize
where the lower diagonal plane partitions are under the condition that the entry $\pi_{N,1}$ is equal to $k$.\\
\indent Finally, considering the \textbf{ascending Young tableaux} interpretation, notice that when we transform from the lower diagonal plane partition to the Young tableau, the fact that $\pi_{N,1}=k$ means that the weight $t_1$ equals $k$. Since the weight $t_1$ does appear, as $v_1$ is not present, we can simply consider the skew partition $\{ \lambda - k\}$ to generate the tableaux, leading to,
\small\begin{equation}\begin{split}
\psi^{(1,k)}_{\{\lambda \}}  = \sum_{T^{\{ \lambda -k \}}_+ }  v^{-2 t_2+M}_2 \dots v^{-2 t_N+M}_N =  (v_2 \dots v_N)^M S_{\{\lambda \}/\{k \}} (v^{-2}_2, \dots, v^{-2}_N)
\end{split}\end{equation}\normalsize
\subsection{Boundary correlation functions - I} Consider then the following boundary correlation function,
\small\begin{equation}
\langle \Psi^{\{k \}}_M|\Psi_M \rangle = \left( \frac{\prod^N_{j=2} v_j}{\prod^N_{j=1} u_j} \right)^M   \sum_{\genfrac{}{}{0mm}{}{\{\lambda\} \subseteq \{(M)^{(N-1)},k\}}{\{\lambda\} \supseteq  \{ k \}}}  S_{\{\lambda \}} ( \{ u^{2}\}) S_{\{\lambda \}/\{k \}} (v^{-2}_2, \dots, v^{-2}_N)
\end{equation}\normalsize
which calculates all the weighted non crossing column strict lattice paths on an $(M+1)\times 2N$ lattice with the first path in the first column turning right at row $k$. Compare it to any of the wave-functions calculated earlier, and concentrate on the particular row, $s=n-M-1 = m+N-1$, of the wave-matrix $\hat{W}^{(0)}$. If we restrict the variables in the usual way (eq. \ref{usual}), and take the $v_1 \rightarrow \infty$ limit we immediately obtain,
\small\begin{equation*}\begin{split}
\lim_{v_1\rightarrow \infty}\tau(n-M-1) \hat{w}^{(0)}_{k}(n-M-1) &=  \sum_{\genfrac{}{}{0mm}{}{\{\lambda\} \subseteq \{(M)^{(N-1)},k\}}{\{\lambda\} \supseteq  \{ k \}}} S_{\{\lambda \}} (\{u^2 \}) S_{\{\lambda \}/\{k \}} (\{v^{-2} \})\\
&= \left(\frac{\prod^N_{j=1} u_j}{\prod^N_{j=2} v_j} \right)^M  \langle \Psi^{\{k \}}_M |\Psi_M \rangle
\end{split}\end{equation*}\normalsize
for $0 \le k \le M$. Thus the wave-vector, given by the $s=n-M-1 = m+N-1$ row of the wave-matrix, in the $v_1 \rightarrow \infty$ limit gives exactly (up to a multiplicative factor) all the weighted non crossing column strict lattice paths on an $(M+1)\times 2N$ lattice with the first path in the first column turning right at row $k$, $0 \le k \le M$.\\
\\
\textbf{Single determinant form for the wave-functions.} When introducing the scalar product, we gave a single determinant form given by eq. \ref{scalardet}. From this expression, it is possible to obtain a single determinant form for the wave-functions considered above\footnote{The details below are given in section VI of \cite{Bog2} to obtain single determinant expressions of boundary 1-point correlation functions. We expand upon these results shortly.}.\\
\indent To achieve this, we first examine the operator $C(v)$ briefly. More explicitly, we are interested in the parts of $C(v)$ that contain only $\phi_j$ operators,
\small\begin{equation}
C(v) = \sum^M_{j=0} v^{M-2j} \phi_j + \textrm{terms that contain operators $\phi^{\dagger}_j$}
\end{equation}\normalsize
Thus when $C(v)$ acts on the conjugate vacuum we obtain,
\small\begin{equation}
\langle 0|C(v) =v^{M} \sum^M_{j=0} v^{-2j} \langle 0| \phi_j 
\end{equation}\normalsize
We use this to express the scalar product as the following weighted linear sum of correlation functions,
\small\begin{equation}
 \langle \Psi_M |\Psi_M \rangle   = v^{M}_1 \sum^M_{j=0} v^{-2j}_1  \langle \Psi^{\{j \}}_M |\Psi_M \rangle 
\end{equation}\normalsize
Therefore, if we expand the single matrix form for the scalar product as a polynomial in $v^2_1$, the coefficients will reveal a single matrix form for the correlation functions/wave-functions.\\
\indent Using the following symmetric polynomial identity,
\small\begin{equation}
h_{p}( \{v^2 \}, v^2_{j})-h_{p}(\{v^2 \},v^2_{k}) =  \left(v^2_{j}-v^2_k  \right) h_{p-1}(\{v^2 \}, v^2_{j},v^2_{k})
\end{equation}\normalsize
where $\{v^2_{j},v^2_{k} \} \notin \{v^2 \}$, we can apply the following row operations,
\small\begin{equation*}
R_{j_k} - R_{N-k+1} \textrm{  ,  } 1 \le j_k \le N-k \textrm{  ,  } 1 \le k \le N-1 
\end{equation*}\normalsize
to completely eliminate $v_1$ from the Vandermonde expression in the scalar product. Additionally, applying the following polynomial expansion of $h_{p}( \{v^2 \}, v^2_{j})$,
\small\begin{equation*}
h_{p}( \{v^2 \}, v^2_{j}) = \sum^p_{q=0} \left(v^2_j \right)^{q} h_{p-q}( \{v^2 \})
\end{equation*}\normalsize
to all entries in the determinant which contain $v^2_1$, we receive,
\small\begin{equation}
 \langle \Psi_M |\Psi_M \rangle =  v^M_1  \sum^M_{q=0} v^{-2q}_1 \Omega_{\hat{v}_1}  \textrm{det}\left[\genfrac{}{}{0mm}{}{h_q(u^2_k,v^2_2,\dots,v^2_N)}{h_{M+N-1}(u^2_k,v^2_j)}\right]_{\genfrac{}{}{0mm}{}{j=2,\dots,N}{k=1,\dots,N }} 
\label{bigom}\end{equation}\normalsize
where,
\small\begin{equation}
 \Omega_{\hat{v}_1} = \prod^N_{i_1 =1}u^{-M}_{i_1}\prod^N_{i_2=2}v^{-M}_{i_2}\prod_{1\le j_1 < k_1 \le N} \frac{1}{u^2_{j_1}-u^2_{k_1}}\prod_{2\le j_2< k_2 \le N} \frac{1}{v^2_{j_2}-v^2_{k_2}}
\label{bigom2}\end{equation}\normalsize
which gives a single determinant form for the (restricted) wave-functions, $\hat{w}^{(0)}$.
\subsection{Wave-functions - II} Considering the $\hat{w}^{(\infty)}$ class of wave-functions, using the definitions given previously we have,
\small\begin{equation*}
\tau(s) \hat{w}^{(\infty)}_k(s) = (-1)^k  \sum_{\{\lambda \} \subseteq (n-s)^{(s-m)}} \chi_{\{\lambda \}/\{1^k \}} (\vec{x}) \chi_{\{\lambda \}} (-\vec{y})
\end{equation*}\normalsize
where we have used the following result,
\small\begin{equation}
\zeta_j (- \tilde{\partial}_{\vec{x}})\chi_{\{\lambda \}} (\vec{x}) = (-1)^j \chi_{\{\lambda \}/ \{1^j \}} (\vec{x})
\end{equation}\normalsize
for all partitions $\{ \lambda \}$ such that $\{ \lambda \} \supseteq \{ 1^j\}$.\\
\indent Thus the lower triangular wave-matrix $\hat{W}^{(\infty)}$ has the form,
\small\begin{equation*}
\hat{W}^{(\infty)}= \left( \frac{(-1)^{j-k}}{\tau(j)}\sum_{\{\lambda \} \subseteq (n-j)^{(j-m)}} \chi_{\{\lambda \}/\{1^{j-k} \}} (\vec{x}) \chi_{\{\lambda \}} (-\vec{y}) \right)_{m \le j,k \le n-1}\\
\end{equation*}\normalsize
\textbf{Constructing $N$-particle state vectors.} Consider the following state vector,
\small\begin{equation*}
 B(u_1)  \dots B(u_{N-k}) \left( \phi^{\dagger}_1 \right)^k |0\rangle = |\Psi^{\{1^k\}}_M \rangle
\end{equation*}\normalsize
The allowable partitions of this vector are given by the following result.
\begin{proposition}
\small\begin{equation}
|\Psi^{\{1^k\}}_M \rangle = \sum_{\genfrac{}{}{0mm}{}{\{\lambda\} \subseteq \{(M)^{(N-k)},1^k\}}{\{\lambda\} \supseteq  \{1^k \}}} \psi^{(2,1^k)}_{\{\lambda \}} |\lambda\rangle
\end{equation}\normalsize
\end{proposition}
\textbf{Proof.} Consider again the lattice path interpretation of the state vectors. The operator(s) $\left( \phi^{\dagger}_1 \right)^k$ assure us that the last $k$ paths, labelled $j_{q}$, $N-k+1 \le q \le N$, make directional changes from right to up at row $1$, column $q$. Thus the largest occupation number sequence can be,
\small\begin{equation*}
\{n_0, n_1, \dots,n_M\} = \{0, k, 0,\dots,0,N-k\} \Rightarrow \{\lambda\} \subseteq \{(M)^{(N-k)},1^k\}
\end{equation*}\normalsize
Also, since columns $\{N-k+1, \dots, N\}$ only contain one $\phi^{\dagger}$ operator each, this means that only columns $\{1, \dots,  N-k\}$ can contain paths in the zeroth row. Due to the paths being column strict  the lowest occupation number sequence is,
\small\begin{equation*}
\{n_0, n_1, \dots,n_M\} = \{N-k, k, 0,\dots,0,0\} \Rightarrow \{\lambda\} \supseteq \{1^k\} \textrm{    } \square
\end{equation*}\normalsize
\textbf{Combinatorial definitions of $\mathbf{\psi^{(2,1^k)}_{\{\lambda \}}}$.} Considering the \textbf{lattice path} interpretation we obtain,
\small\begin{equation*}
\psi^{(2,1^k)}_{\{\lambda \}}= \sum_{\genfrac{}{}{0mm}{}{\textrm{allowable paths in}}{\textrm{$(M+1)\times N$ lattice$^\dagger$}}} u^{t^{d}_1-t^{a}_1}_1 \dots u^{t^{d}_{N-k}-t^{a}_{N-k}}_{N-k}
\end{equation*}\normalsize
where the lattice paths are under the condition that the last $k$ paths, labelled $j_{q}$, $N-k+1 \le q \le N$, make directional changes from right to up at row $1$, column $q$, and only columns $\{1, \dots,  N-k\}$ can contain paths in the zeroth row. \\
\indent Considering the \textbf{upper diagonal plane partition} interpretation we obtain,
\small\begin{equation*}
\psi^{(2,1^k)}_{\{\lambda \}}= \sum_{\genfrac{}{}{0mm}{}{\textrm{upper diag. plane part.}}{\textrm{in $N \times N \times M$ array$^\dagger$}}} u^{l^{d}_1-l^{a}_1}_1 \dots u^{l^{d}_{N-k}-l^{a}_{N-k}}_{N-k}
\end{equation*}\normalsize
where the upper plane partitions are under the condition that the top-right most $k \times k$ entries are equal to one. This obviously places restrictions on the remaining entries, as per the conditions of a plane partition. For example, the remaining $(N-k)\times k$ bottom-right entries can only either be zero or one accordingly.\\
\indent Finally, whenever we biject from the upper plane partitions to the \textbf{descending Young tableaux}, the weights $t_{N-k+1}= \dots = t_{N} = 1$, and their position in the tableau are exactly $\{t_N = T^{\{\lambda\}}_{1,1},t_{N-1} = T^{\{\lambda\}}_{2,1},\dots,t_{N-k+1} = T^{\{\lambda\}}_{k,1}  \}$. Since these weights do not enter the equation, due to $u_{N-k+1},\dots,u_N$ not being present, we can consider the skew partition $\{\lambda - 1^k\}$ to generate the tableaux\footnote{Incidentally, it is at this point the reason we considered the tableaux in descending order becomes apparent. Had we considered ascending order we would need to invert the numbers to obtain the required results.}. Thus we obtain,
\small\begin{equation}
\psi^{(2,1^k)}_{\{\lambda \}}  = \sum_{T^{\{ \lambda -1^k \}}_- }  u^{2 t_1-M}_1 \dots u^{2 t_{N-k}-M}_{N-k} =   \left(\frac{1}{u_1 \dots u_{N-k}}\right)^M  S_{\{\lambda \}/\{1^k \}} (u^{2}_1, \dots, u^{2}_{N-k})
\end{equation}\normalsize
\subsection{Boundary correlation functions - II} Consider then the boundary correlation function,
\small\begin{equation}
 \langle\Psi_M |\Psi^{\{1^k\}}_M \rangle =\left( \frac{\prod^{N}_{j=1} v_j}{\prod^{N-k}_{j=1} u_j} \right)^M \sum_{\genfrac{}{}{0mm}{}{\{\lambda\} \subseteq \{(M)^{(N-k)},1^k\}}{\{\lambda \} \supseteq  \{1^k\}}}  S_{\{\lambda \}/\{1^k \}} (u^{2}_1, \dots, u^{2}_{N-k}) S_{\{\lambda \}} (\{v^{-2}\})
\end{equation}\normalsize
which calculates all the weighted non crossing column strict lattice paths on an $(M+1)\times 2N$ lattice with the final $k$ paths, labelled $j_q$, $N-k+1 \le q \le N$, turning up at row 1, column $N-k+1 \le q \le N$. Additionally, only columns $1 \le q \le N-k$ can contain paths in the zeroth row. Compare the above result with the $s=n-M = m+N$ row of the wave-matrix $\hat{W}^{(\infty)}$, restricting the variables as usual, and taking the $u_{N-k+1} = \dots = u_N = 0$ limit,
\small\begin{equation}\begin{split}
\lim_{ \genfrac{}{}{0mm}{}{u_{j} \rightarrow 0 }{N-k+1 \le j \le N} }  \tau(n-M) \hat{w}^{(\infty)}_{k}(n-M) &= (-1)^k \sum_{\genfrac{}{}{0mm}{}{\{\lambda\} \subseteq \{M^{(N-k)},1^k\}}{\{\lambda\} \supseteq  \{1^k \}}} S_{\{\lambda \}/\{1^k \}} (\{ u^2\}) S_{\{\lambda \}} (\{v^{-2}\})\\
&=(-1)^k  \left( \frac{\prod^{N-k}_{j=1} u_j}{\prod^N_{j=1} v_j} \right)^M   \langle \Psi_M |\Psi^{\{1^k\}}_M \rangle 
\end{split}\end{equation}\normalsize
for $0 \le k \le M$. Thus the wave-vector, given by the $s=n-M = m+N$ row of the wave-matrix, in the $u_{N-k+1} = \dots = u_N = 0$ limit gives exactly (up to a multiplicative factor) all the weighted non crossing column strict lattice paths on an $(M+1)\times 2N$ lattice with the final $k$ paths, $1< k \le N$, labelled $j_q$, $N-k+1 \le q \le N$, turning up at row 1, column $N-k+1 \le q \le N$ and only the first $N-k$ columns can contain paths in the zeroth row.\\
\\
\textbf{Single determinant form for the wave-functions.} We begin by examining the operator $B(u)$, as we are interested in the parts of $B(u)$ that contain only $\phi^{\dagger}_j$ and $\phi_1$ operators,
\small\begin{equation}\begin{split}
B(u) &=u^{-M} \left\{ \sum^M_{j=0} u^{2j} \phi^{\dagger}_j +\sum^{M-2}_{j=0} u^{2j+2} \phi^{\dagger}_0 \phi_1 \phi^{\dagger}_{j+2} \right\} \\
& \textrm{$+$ terms that contain operators $\phi_j$   ,   $j \in \{2,\dots,M\}$} 
\end{split}\label{BB}\end{equation}\normalsize
\textbf{1-point boundary functions.} Following the corresponding workings from section 3.2, we can obtain the scalar product as the following weighted linear sum of 1-point boundary correlation functions,
\small\begin{equation}
 \langle\Psi_M |\Psi_M \rangle   = u^{-M}_N \sum^M_{j=0} u^{2j}_N  \langle\Psi_M |\Psi^{\{j\}}_M \rangle
\end{equation}\normalsize
Explicitly expanding the determinant expression as a polynomial in $u^2_{N}$ we obtain,
\small\begin{equation}\begin{split}
\langle\Psi_M |\Psi^{\{q\}}_M \rangle = \Omega_{\hat{u}_N} \textrm{det} \left[ h_{M+N-k}(\{u^2\}_{k},v^2_j), h_{M-q}(\{u^2\}_{N-1},v^2_j)\right]_{\genfrac{}{}{0mm}{}{j=1,\dots,N}{k=1,\dots,N-1 }}
\label{nice33}\end{split}\end{equation}\normalsize
where $\{u^2\}_{k} = \{u^2_1, \dots, u^2_k  \}$ and $\Omega_{\hat{u}_N}$ is the equivalent expression of eq. \ref{bigom2}.\\
\\
\textbf{2-point boundary functions.} We now build upon eq. \ref{nice33} and use eq. \ref{BB} to consider the following quantity,
\small\begin{equation}
B(u_{N-1})\phi^{\dagger}_1|0\rangle =u^{-M}_{N-1} \sum^M_{j=0} u^{2j}_{N-1}  \phi^{\dagger}_j \phi^{\dagger}_1 |0\rangle + u^{-(M+2)}_{N-1} \sum^{M-2}_{j=2} u^{2j}_{N-1}  \phi^{\dagger}_0  \phi^{\dagger}_{j} |0\rangle
\end{equation}\normalsize
Hence we can express the 1-point correlation  function $\langle\Psi_M |\Psi^{\{1\}}_M \rangle$, as the following linear sum of 2-point correlation functions,
\small\begin{equation}\begin{split}
\langle\Psi_M |\Psi^{\{1\}}_M \rangle &=  u^{-M}_{N-1}\left\{ \langle\Psi_M |\Psi^{\{1,0\}}_M \rangle + u^{2M}_{N-1} \langle\Psi_M |\Psi^{\{M,1\}}_M \rangle \right\}   \\
&+  \textrm{   } u^{-M}_{N-1} \sum^{M-1}_{j=1} u^{2j}_{N-1}\left\{ \langle\Psi_M |\Psi^{\{j,1\}}_M \rangle + \langle\Psi_M |\Psi^{\{j+1,0\}}_M \rangle   \right\}
\end{split}\end{equation}\normalsize
where the coefficient of $u^{-M+2}_{N-1}$ is $\langle\Psi_M |\Psi^{\{1^2\}}_M \rangle + \langle\Psi_M |\Psi^{\{2,0\}}_M \rangle $. Thus in order to obtain $\langle\Psi_M |\Psi^{\{1^2\}}_M \rangle$, we need to first find $\langle\Psi_M |\Psi^{\{2,0\}}_M \rangle $.\\
\indent This can be achieved by expanding $\langle\Psi_M |\Psi^{\{0\}}_M \rangle$ as a series in $u^2_{N-1}$,
\small\begin{equation*}
\langle\Psi_M |\Psi^{\{0\}}_M \rangle = u^{-M}_{N-1} \sum^M_{j=0}u^{2j}_{N-1} \langle\Psi_M |\Psi^{\{j,0\}}_M \rangle
\end{equation*}\normalsize
Substituting $q=0$ into eq. \ref{nice33} and expanding as a polynomial in $u^{2}_{N-1}$ we obtain,
\small\begin{equation}\begin{split}
 \langle\Psi_M |\Psi^{\{q,0\}}_M \rangle = \Omega_{\hat{u}_{N},\hat{u}_{N-1}} \textrm{det} \left[ c_{jk},h_{M+1}(\{u^2\}_{N-2},v^2_j),h_{M-q}(\{u^2\}_{N-2},v^2_j) \right]_{\genfrac{}{}{0mm}{}{j=1, \dots, N}{k=1,\dots,N-2 }}
\label{already}\end{split}\end{equation}\normalsize
where $c_{jk} = h_{M+N-k}(\{u^2\}_{k},v^2_j)$.\\
\indent With the above result, we now expand the determinant form of $\langle\Psi_M |\Psi^{\{1\}}_M \rangle$, (the $q=1$ case of eq. \ref{nice33}), as a polynomial in $u^{2}_{N-1}$,
\small\begin{equation*}\begin{split}
\langle\Psi_M |\Psi^{\{1\}}_M \rangle &=u^{-M}_{N-1} \sum^{M+1}_{q_1=2}\sum^{1}_{q_2=0}u^{2q_1+2 q_2-4}_{N-1}\Omega_{\hat{u}_{N},\hat{u}_{N-1}}\\
& \times \textrm{det}\left[ c_{jk},  h_{M-q_2+1}(\{u^2\}_{N-2},v^2_{j}),h_{M-q_1+1}(\{u^2\}_{N-2},v^2_{j}) \right]_{\genfrac{}{}{0mm}{}{j=1,\dots,N}{k=1,\dots,N-2 }}
\end{split}\end{equation*}\normalsize
where the indices $(q_1,q_2) = (q,1)$, and $(q+1,0)$, $2 \le q \le M$, give us the sum $\langle\Psi_M |\Psi^{\{q-1,1\}}_M \rangle + \langle\Psi_M |\Psi^{\{q,0\}}_M \rangle$. Since we already have the explicit form of $\langle\Psi_M |\Psi^{\{q,0\}}_M \rangle$, given in eq. \ref{already}, this leaves us with the required result,
\small\begin{equation}
\langle\Psi_M |\Psi^{\{q,1\}}_M \rangle = \Omega_{\hat{u}_{N},\hat{u}_{N-1}} \textrm{det}\left[ c_{jk},  h_{M}(\{u^2\}_{N-2},v^2_{j}), h_{M-q}(\{u^2\}_{N-2},v^2_{j}) \right]_{\genfrac{}{}{0mm}{}{j=1,\dots,N}{k=1,\dots,N-2 }}
\end{equation}\normalsize
\textbf{n-point boundary functions.} Given the previous examples, we present the following result.
\begin{proposition}
\small\begin{equation}\begin{split}
\langle\Psi_M |\Psi^{\{r_1,\dots,r_n\}}_M \rangle &= \Omega_{\hat{u}_{N},\dots,\hat{u}_{N+1-n}} \textrm{det}\left[ c_{jk},  h_{M-r_n+n-1}(\{u^2\}_{N-n},v^2_{j}),\right. \\
&\left. h_{M-r_{n-1}+n-2}(\{u^2\}_{N-n},v^2_{j}),\dots,h_{M-r_1}(\{u^2\}_{N-n},v^2_{j}) \right]_{\genfrac{}{}{0mm}{}{j=1,\dots,N}{k=1,\dots,N-n }}
\end{split}\label{tochi}\end{equation}\normalsize
where,
\small\begin{equation*}\begin{split}
r_1 \in \{ 0,1, \dots,M \} \textrm{  ,  } r_2 \in \{0,1\} \textrm{  , \dots ,  } r_n \in \{ 0,1\}\\
r_1 \ge  r_2 \ge \dots \ge r_n \textrm{   ,   } 1 \le n \le N
\end{split}\end{equation*}\normalsize
\end{proposition}
\textbf{Comment.} The above statement can be proven using induction. By assuming that eq. \ref{tochi} is true, we obtain $\langle\Psi_M |\Psi^{\{1^{n-q},0^{q} \}}_M \rangle$, $0 \le q \le n-1$, as the following weighted sum of $(n+1)$-point correlation functions,
\small\begin{equation}\begin{split}
\langle\Psi_M |\Psi^{\{1^{n-q},0^{q} \}}_M \rangle &= u^{-M}_{N-n} \langle\Psi_M |\Psi^{\{1^{n-q},0^{q+1} \}}_M \rangle + u^{M}_{N-n} \langle\Psi_M |\Psi^{\{M,1^{n-q},0^{q} \}}_M \rangle +u^{-M}_{N-n}  \\
&\times  \sum^{M-1}_{j=1} u^{2j}_{N-n}\left\{ \langle\Psi_M |\Psi^{\{j,1^{n-q},0^{q}  \}}_M \rangle + \langle\Psi_M |\Psi^{\{j+1,1^{n-q-1},0^{q+1} \}}_M \rangle   \right\}
\label{machma}\end{split}\end{equation}\normalsize
Additionally for $\langle\Psi_M |\Psi^{\{0^{n} \}}_M \rangle$, $(q=n)$, we have,
\small\begin{equation}
\langle\Psi_M |\Psi^{\{0^{n} \}}_M \rangle=u^{-M}_{N-n}\sum^M_{j=0} u^{2j}_{N-n} \langle\Psi_M |\Psi^{\{j,0^{n} \}}_M \rangle 
\label{ache2}\end{equation}\normalsize
In order to verify the proposed result we need to derive, using polynomial expansion method(s) on eq. \ref{tochi}, the explicit determinant forms for the following expressions (referred to as step $\mathbf{1,2}$ and $\mathbf{3}$),
\begin{itemize}
\item{\textbf{Step 1)} $\langle\Psi_M |\Psi^{\{j,0^{n} \}}_M \rangle$, the coefficient of $u^{-M+2j}_{N-n}$ in eq. \ref{ache2}.}
\item{\textbf{Step 2)} $\langle\Psi_M |\Psi^{\{1^{n-q},0^{q+1} \}}_M \rangle$ and $\langle\Psi_M |\Psi^{\{M,1^{n-q},0^{q} \}}_M \rangle$, the coefficients of $u^{-M}_{N-n}$ and $u^{M}_{N-n}$ in eq. \ref{machma}.}
\item{\textbf{Step 3)} $\langle\Psi_M |\Psi^{\{j,1^{n-q},0^{q} \}}_M \rangle$ and $\langle\Psi_M |\Psi^{\{j+1,1^{n-q-1},0^{q+1} \}}_M \rangle$,  the coefficients of $u^{-M+2j}_{N-n}$ in eq. \ref{machma}.}
\end{itemize}
It is necessary in the third step to determine which of the determinant expressions corresponds to which correlation function. This is achieved through considering values of $q$ where one term is already known from a previous result. As an example, consider $q=n-1$, where we receive the sum $\langle\Psi_M |\Psi^{\{j,1,0^{n-1} \}}_M \rangle + \langle\Psi_M |\Psi^{\{j+1,0^{n} \}}_M \rangle$. In this case we have already obtained $\langle\Psi_M |\Psi^{\{j+1,0^{n} \}}_M \rangle$ from the first step. For $q=n-2$ we receive the sum $\langle\Psi_M |\Psi^{\{j,1^2,0^{n-2} \}}_M \rangle + \langle\Psi_M |\Psi^{\{j+1,1,0^{n-1} \}}_M \rangle$, where we know the value of $\langle\Psi_M |\Psi^{\{j+1,1,0^{n-1} \}}_M \rangle$, $1 \le j \le M-2$, from the previous $(q=n-1)$ calculation, and we know $j = M-1$ from the second step. Carefully following this argument for all values of $q$ we complete the proof by induction.\\
\indent Thus using the results of eq. \ref{tochi}, we obtain the single determinant form for the restricted wave-functions, $\hat{w}^{(\infty)}$.
\section{Discussion}
The main result of this work is the correspondence between the 2-Toda wave-functions and the boundary correlation functions of the phase model. The weighted sum of the wave-functions analyzed in this work can be thought of as the action of a single vertex operator \cite{bluebook} on the 2-Toda $\tau$-function,
\small\begin{equation*}
\frac{\Gamma_x(\lambda) \tau(s,\vec{x},\vec{y} )}{\tau(s,\vec{x},\vec{y} )}=  \sum^{s-m}_{k=0} \lambda^k \hat{w}^{(\infty)}_{k}(s) \textrm{  ,  } \frac{\Gamma_y(\lambda) \tau(s+1,\vec{x},\vec{y} )}{\tau(s,\vec{x},\vec{y} )}=  \sum^{n-s-1}_{k=0} \lambda^k \hat{w}^{(0)}_{k}(s)
\end{equation*}\normalsize
where $\Gamma_{x/y}(\lambda)=\exp \left\{ - \sum^{n-m-1}_{k=1} \frac{\lambda^k}{k} \partial_{x_k}/\partial_{y_k} \right\}$. It is a pertinent question as to whether there is a combinatorial interpretation of a $\tau$-function that has been acted on by more than one vertex operator. Alternatively, it is perfectly natural to speculate whether \textit{non boundary} correlation functions of the phase model have natural correspondences with fundamental objects of the 2-Toda hierarchy. It is the author's intentions to examine these and related questions in the future.
\section*{Acknowledgements}
The author would like to thank O~Foda for discussions, and M~Wheeler for making him aware of the plane partition/tableau bijection, without which this work would not exist. This work was supported by the Dept. of Mathematics and Statistics, The University of Melbourne.

\end{document}